\documentclass[fleqn,usenatbib]{mnras}

\usepackage{newtxtext,newtxmath}

\usepackage[T1]{fontenc}

\DeclareRobustCommand{\VAN}[3]{#2}
\let\VANthebibliography\thebibliography
\def\thebibliography{\DeclareRobustCommand{\VAN}[3]{##3}\VANthebibliography}

\usepackage{graphicx}	
\usepackage{amsmath}	
\usepackage{xcolor}
\usepackage{caption}
\usepackage{subcaption}
\usepackage{lineno}
\usepackage{academicons}
\usepackage{eso-pic}

\AddToShipoutPictureBG*{%
  \AtPageUpperLeft{%
    \hspace{0.75\paperwidth}%
    \raisebox{-3.5\baselineskip}{%
      \makebox[0pt][l]{\textnormal{DES-2025-0895}}
}}}%

\AddToShipoutPictureBG*{%
  \AtPageUpperLeft{%
    \hspace{0.75\paperwidth}%
    \raisebox{-4.5\baselineskip}{%
      \makebox[0pt][l]{\textnormal{FERMILAB-PUB-25-0594-PPD}}
}}}%

\definecolor{ork}{rgb}{0.9,0.1,0.3}
\definecolor{grbl}{rgb}{0.3,0.6,0.7}
\definecolor{bleu}{rgb}{0,0.5,0.6}
\definecolor{mypink3}{cmyk}{0, 0.7808, 0.4429, 0.1412}

\newcommand{\cosmoparams}{\ensuremath{\boldsymbol{\eta}}}
\newcommand{\orcid}[1]{\href{https://orcid.org/#1}{\textcolor[HTML]{A6CE39}{\aiOrcid}}}

\def\mpcoh{{\,h^{-1}\,\rm Mpc}}

\DeclareMathOperator{\Avg}{Avg}
\DeclareMathOperator{\RMS}{RMS}

\let\oldequation\equation
\let\oldendequation\endequation
\renewenvironment{equation}
  {\linenomathNonumbers\oldequation}
  {\oldendequation\endlinenomath}

\title[Galaxy biasing of troughs and peaks]{Biasing from galaxy trough and peak profiles with the DES Y3 redMaGiC galaxies and the weak lensing mass map}

\author[DES Collaboration]{
\parbox{\textwidth}{
\Large
Q.~Hang,$^{1}$\thanks{\url{e.hang@ucl.ac.uk}}
N.~Jeffrey,$^{1}$
L.~Whiteway,$^{1}$
O.~Lahav,$^{1}$
J.~Williamson,$^{1}$
M.~Gatti,$^{2}$
J.~DeRose,$^{3}$
A.~Kovacs,$^{4,5}$
A.~Alarcon,$^{6}$
A.~Amon,$^{7}$
K.~Bechtol,$^{8}$
M.~R.~Becker,$^{9}$
G.~M.~Bernstein,$^{10}$
A.~Campos,$^{11,12}$
A.~Carnero~Rosell,$^{13,14,15}$
M.~Carrasco~Kind,$^{16,17}$
C.~Chang,$^{18,2}$
R.~Chen,$^{19}$
A.~Choi,$^{20}$
S.~Dodelson,$^{18,21,2}$
C.~Doux,$^{10,22}$
A.~Drlica-Wagner,$^{18,21,2}$
J.~Elvin-Poole,$^{23}$
S.~Everett,$^{24}$
A.~Fert\'e,$^{25}$
D.~Gruen,$^{26}$
R.~A.~Gruendl,$^{16,17}$
I.~Harrison,$^{27}$
M.~Jarvis,$^{10}$
N.~MacCrann,$^{28}$
J.~McCullough,$^{7,29,25,26}$
J.~Myles,$^{7}$
A. Navarro-Alsina,$^{30}$
S.~Pandey,$^{10}$
J.~Prat,$^{18,31}$
M.~Raveri,$^{32}$
R.~P.~Rollins,$^{33}$
E.~S.~Rykoff,$^{29,25}$
C.~S{\'a}nchez,$^{10}$
L.~F.~Secco,$^{2}$
I.~Sevilla-Noarbe,$^{34}$
E.~Sheldon,$^{35}$
T.~Shin,$^{36}$
M.~A.~Troxel,$^{19}$
I.~Tutusaus,$^{37}$
R.~H.~Wechsler,$^{38,29,25}$
B.~Yanny,$^{21}$
B.~Yin,$^{19}$
M.~Aguena,$^{39,14}$
O.~Alves,$^{40}$
F.~Andrade-Oliveira,$^{41}$
D.~Bacon,$^{42}$
J.~Blazek,$^{43}$
S.~Bocquet,$^{26}$
D.~Brooks,$^{1}$
J.~Carretero,$^{44}$
R.~Cawthon,$^{45}$
M.~Crocce,$^{46,6}$
L.~N.~da Costa,$^{14}$
M.~E.~da Silva Pereira,$^{47}$
T.~M.~Davis,$^{48}$
S.~Desai,$^{49}$
H.~T.~Diehl,$^{21}$
P.~Doel,$^{1}$
B.~Flaugher,$^{21}$
J.~Frieman,$^{18,21,2}$
G.~Gutierrez,$^{21}$
S.~R.~Hinton,$^{48}$
D.~L.~Hollowood,$^{50}$
K.~Honscheid,$^{51,52}$
K.~Kuehn,$^{53,54}$
S.~Lee,$^{55}$
J.~L.~Marshall,$^{56}$
J. Mena-Fern{\'a}ndez,$^{22}$
R.~Miquel,$^{57,44}$
A.~A.~Plazas~Malag\'on,$^{29,25}$
A.~Porredon,$^{34,58}$
A.~Roodman,$^{25,29}$
S.~Samuroff,$^{43,44}$
E.~Sanchez,$^{34}$
D.~Sanchez Cid,$^{34,41}$
M.~Smith,$^{59}$
E.~Suchyta,$^{60}$
M.~E.~C.~Swanson,$^{16}$
C.~To,$^{18}$
and V.~Vikram$^{10}$
\begin{center} (DES Collaboration) \end{center}
}
\vspace{0.4cm}
\\
\parbox{\textwidth}{
Affiliations are listed at the end of the paper.
}
}

\date{Accepted XXX. Received YYY; in original form ZZZ}

\pubyear{2025}

\begin{document}
\label{firstpage}
\pagerange{\pageref{firstpage}--\pageref{lastpage}}
\maketitle

\begin{abstract}
We measure the correspondence between the distribution of galaxies and matter around troughs and peaks in the projected galaxy density, by comparing \texttt{redMaGiC} galaxies ($0.15<z<0.65$) to weak lensing mass maps from the Dark Energy Survey (DES) Y3 data release. We obtain stacked profiles, as a function of angle $\theta$, of the galaxy density contrast $\delta_{\rm g}$ and the weak lensing convergence $\kappa$, in the vicinity of these identified troughs and peaks, referred to as `void' and `cluster' superstructures. The ratio of the profiles depend mildly on $\theta$, indicating good consistency between the profile shapes. We model the amplitude of this ratio using a function $F(\cosmoparams, \theta)$ that depends on cosmological parameters $\cosmoparams$, scaled by the galaxy bias. We construct templates of $F(\cosmoparams, \theta)$ using a suite of $N$-body (`Gower Street') simulations forward-modelled with DES Y3-like noise and systematics. We discuss and quantify the caveats of using a linear bias model to create galaxy maps from the simulation dark matter shells. We measure the galaxy bias in three lens tomographic bins (near to far): $2.32^{+0.86}_{-0.27}, 2.18^{+0.86}_{-0.23}, 1.86^{+0.82}_{-0.23}$ for voids, and $2.46^{+0.73}_{-0.27}, 3.55^{+0.96}_{-0.55}, 4.27^{+0.36}_{-1.14}$ for clusters, assuming the best-fit \textit{Planck} cosmology. Similar values with $\sim0.1\sigma$ shifts are obtained assuming the mean DES Y3 cosmology. The biases from troughs and peaks are broadly consistent, although a larger bias is derived for peaks, which is also larger than those measured from the DES Y3 $3\times2$-point analysis. This method shows an interesting avenue for measuring field-level bias that can be applied to future lensing surveys.
\end{abstract}

\begin{keywords}
gravitational lensing: weak -- large-scale structure of Universe -- cosmology: observations
\end{keywords}





\section{Introduction}

Galaxies are biased tracers of the underlying dark matter density field. It has been known since early galaxy surveys that clusters have higher two-point correlation amplitudes than galaxies \citep{1983ApJ...270...20B}. The difference can be attributed to the amplification of the two-point function of a Gaussian random field that arises when the field is sampled only above a threshold {in the field value} \citep{1984ApJ...284L...9K,1986ApJ...304...15B}. Later, it was noted that such an amplification, or galaxy bias $b$, was needed to reconcile the galaxy two-point function $\xi_{\rm g}(r)$ from galaxy clustering measurements on large scales with the theoretical matter power spectrum $\xi_{\rm m}(r)$ from the cold dark matter (CDM) paradigm: $\xi_{\rm g}(r)=b^2\xi_{\rm m}(r)$ \citep{1985ApJ...292..371D}. 
There is a corresponding relation at the field level {(given sufficient smoothing of the field)}: if $\delta_{\rm m}$ is the matter density contrast and $\delta_{\rm g}(\mathbf{x})$ is the galaxy density contrast (so that the Poisson-distributed galaxy count $N$ in a small volume around $\mathbf{x}$ will have an expected value that is $\langle N \rangle = [1 + \delta_{\rm g}(\mathbf{x})]\bar{N}$, where $\bar{N}$ is the global average of $N$) then $1 + \delta_{\rm g}(\mathbf{x}) = 1 + b \ \delta_{\rm m}(\mathbf{x})$\footnote{Note that this quantity must be non-negative; the consequences of this will be discussed later.}.

The assumption of a constant, {non-stochastic}, and linear $b$ holds only when both fields are close to zero. 
Using haloes in $N$-body simulations, \cite{1999ApJ...520...24D,2005MNRAS.356..247W,Manera_2011} showed that the joint distribution between $\delta_{\rm h}$, the halo density contrast, and $\delta_{\rm m}$ at high and low densities (red and blue galaxies) deviates from linearity with different stochasticity. 
{These stochasticities encode information such as galaxy assembly history. The distribution of $\delta_{\rm m}$ itself also encodes non-Gaussian information about non-linear structure formation at the tails, which can in turn affect galaxy formation in these specific environments \citep{Jing_2007,Gao_2007}. 
These features make troughs and peaks in the density field interesting, providing information complementary to 2-point statistics \citep[e.g.][]{Pelliciari_2023}. }

In this paper, we investigate the correspondence between {the distribution of galaxies} and matter in the troughs and peaks of the projected 2D galaxy density field and measure the galaxy bias in these regions.
The projected dark matter density fluctuations can be probed by weak gravitational lensing through small distortion of background galaxy shapes, referred to as `shear'. 
Throughout this paper, we refer to foreground \textit{lens} and background \textit{source} galaxy samples, where the matter traced by the former acts as gravitational lenses that distort the shapes of the galaxies in the latter.
The galaxy bias can therefore be measured with the 2-point {cross-correlation} function between the lens galaxy and shear. This can be done using a fixed cosmology\footnote{{The bias can be highly degenerate with cosmological parameters, hence fixing (at a wrong) cosmology has the potential to yield incorrect results. This will not be a problem if the uncertainty of cosmological parameters is much smaller than the error on the bias parameter.}} \citep[e.g.][]{2018MNRAS.473.1667P}, or else jointly with cosmological parameters -- and with the addition of galaxy clustering and cosmic shear information -- in the so-called $3\times2$-point analysis \citep[e.g.][]{2022PhRvD.105b3520A}; in the latter the combination with the other statistics breaks the degeneracy between galaxy bias and the amplitude of clustering, $\sigma_8$ (defined as the RMS of the linear matter density field in spheres of radius $8 h^{-1}{\rm Mpc}$ at redshift zero).
Galaxy bias can also be measured for galaxy troughs only. \cite{2016MNRAS.455.3367G} measured the shear signal from cut-outs of low density regions on the projected galaxy fields from the Dark Energy Survey (DES) Science Verification (SV) data, also referred to as the `lensing-in-cell' approach. {When fitting the theory to their measurements, they also found that the results are insensitive to the galaxy bias used in the fit.}
\cite{2018PhRvD..98b3508F} and \cite{2018PhRvD..98b3507G} combined this method with the galaxy `counts-in-cell' and extended the measurements to five galaxy density quantiles, using the DES Year 1 \texttt{redMaGiC} and Sloan Digital Sky Survey (SDSS) galaxy catalogues.
The measurements were used to constrain stochastic galaxy biasing models, along with cosmological parameters.
These studies show that weak lensing is a powerful tool for studying the connection between galaxies and dark matter.

A more direct comparison between light and matter can be achieved using weak lensing \textit{mass maps} i.e. maps of the lensing convergence $\kappa$; this is because $\kappa$ is directly proportional to the line-of-sight projected dark matter density contrast (subject to a distance-dependent kernel). Mass maps can be reconstructed from the measured shear map.
Such a comparison was done for DES data with various reconstruction methods \citep{2015PhRvL.115e1301C,Chang_2018,2021MNRAS.505.4626J}.
The galaxy bias can be measured at the map level using the galaxy density and mass maps. For example, \cite{2012MNRAS.424..553A} and \cite{2016MNRAS.462...35P} constructed the zero-lag correlation function between $\kappa$ and the bias-weighted convergence field, $\kappa_{\rm g}$, and showed that the galaxy bias can be directly measured by taking the ratio $\langle \kappa_{\rm g} \kappa \rangle /\langle \kappa \kappa \rangle$. \cite{2016MNRAS.459.3203C} applied this method to the DES SV data to measure how the linear galaxy bias evolved with redshift. Unlike analyses using galaxy clustering and galaxy-galaxy lensing, these bias measurements are independent of $\sigma_8$ values.
A challenge of using mass maps in the analysis is to account for noise and for systematic effects (such as survey masks, redistribution of matter due to baryonic physics, and galaxy intrinsic alignments) that can degrade and bias the reconstructed convergence; this can be tackled by simulation-assisted forward modelling.

We take an approach similar to that of \cite{2016MNRAS.459.3203C}, but with a focus on galaxy density troughs and peaks. We measure the galaxy density and $\kappa$ profiles as a function of angular scales $\theta$ around troughs and peaks identified in the galaxy field, using the DES Y3 \texttt{redMaGiC} galaxies and mass maps. By comparing the profile shapes via their ratios, $\langle \kappa (\theta) \rangle/\langle \delta_{\rm g} (\theta)\rangle$, we quantify the galaxy bias given a fixed cosmology.
As a first step, we shall focus on linear scales.
We use the \textit{Gower Street} simulations \citep{2025MNRAS.536.1303J}, a suite of $N$-body simulations, to create mock catalogues and forward model systematics, noises, and mask effects, and isolate the dependence of the profile ratio on the lensing kernel and cosmology.


The paper is organized as follows. 
In Section~\ref{sec: data} we introduce the dataset and simulations used in this study. 
In Section~\ref{sec: methods} we outline our approach to identifying density troughs and peaks, discuss different galaxy biases in the Gower Street simulations, and present the profile measurements.
In Section~\ref{sec: model} we detail the forward model approach using a simulation-based template to quantify the ratio between the profiles of {the distribution of galaxies} and matter, and we define the likelihood and covariance matrix used for the analysis. We also validate our pipeline on the Gower Street simulations. The results are presented in Section~\ref{sec: Results}. Finally, we summarize in Section~\ref{sec: conclusions} and provide an outlook on possible future investigations.

\section{Data}
\label{sec: data}

\begin{figure}
	\includegraphics[width=\linewidth]{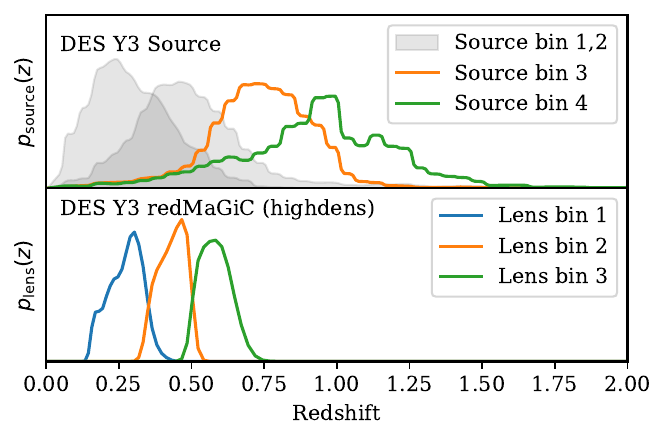}
    \caption{Redshift distribution of the DES Y3 source sample (upper panel) and the high-density \texttt{redMaGiC} galaxies (lower panel). The source redshift distributions are adopted from the mean of \textsc{HyperRank} \citep{2022MNRAS.511.2170C}. For the lens redshift distributions, we show the average distribution combining four realizations of the photometric redshift for each object. Source bins 1 and 2 are not used in the analysis due to their redshift overlap with the lens bins.}
    \label{fig: Buzzard_redmagic_sim_nz}
\end{figure}

\subsection{DES Y3 redMaGiC galaxies}
\label{sec: redmagic} 

DES Y3 is the catalogs and observations from the first 3 years of DES data-taking. It is publicly released as DES Data Release 1 \citep[DR1; ][]{2018ApJS..239...18A}.
The DES Y3 \texttt{redMaGiC} catalogue consists of Luminous Red Galaxies (LRGs) selected using the \texttt{redMaGiC} algorithm \citep{2016MNRAS.461.1431R}. The algorithm works by fitting those galaxies that are above a luminosity threshold $L_*$ to a template of the magnitude-colour-redshift relation. Galaxies are selected if they have a $\chi^2$ goodness of fit smaller than a threshold value $\chi_{\rm max}^2$. 
The \texttt{redMaGiC} catalogue is further split into two subsamples: a lower-redshift `high-density' sample with $L_{\rm min}>0.5L_*$ and a higher-redshift `high-luminosity' sample with $L_{\rm min}>1.0L_*$. We do not use the `high-luminosity' sample due to its overlap with the source redshift distributions.
Each galaxy is assigned a weight $\omega$, inversely proportional to the angular selection function, to account for correlations with survey properties \citep{2022MNRAS.511.2665R}. 

The photometric redshifts (photo-$z$) of \texttt{redMaGiC} galaxies are determined and calibrated in \citet{2016MNRAS.461.1431R} and \cite{2022MNRAS.513.5517C}, with a precision of $\sigma_z/(1+z)\lesssim0.0126$. 
The catalogue provides, for each galaxy, a photometric redshift point estimate (\texttt{zredmagic}), and four additional photo-$z$ realizations drawn from the galaxy's redshift probability distribution. 
We split the lens tomographic bins using the point estimate redshifts, compute the ensemble redshift distribution, $p_{\rm lens}(z)$, using the histograms of the four realizations, and take the average. 
The redshift distribution is normalized such that $\int\, dz\, p_{\rm lens}(z) = 1$.
In this way, we incorporate the redshift uncertainty in the lens sample.
We adopt the same lens tomographic binning as in the DES Y3 $3\times2$-point analysis \citep{2022PhRvD.105b3520A}, i.e. $0.15<z\leq0.35, 0.35<z\leq 0.50, 0.50 <z\leq0.65$. 
The lower panel of Fig.~\ref{fig: Buzzard_redmagic_sim_nz} shows the weighted lens redshift distribution in the three tomographic bins.

The galaxies in each tomographic bin are further binned on the sky into a HEALPix\footnote{\url{http://healpix.sf.net}} \citep[][]{ 2005ApJ...622..759G,Zonca2019} map with ${\rm nside}=512$.
Each pixel has a survey completeness factor $f_{\rm mask}$ determined by the mask \citep{Hartley_2021}; the pixel is masked if $f_{\rm mask} = 0$ and the survey footprint consists of those pixels where $f_{\rm mask} > 0$.
The effective number of galaxies in a pixel is $\hat{n}_{\rm gal} = {\sum \omega}/f_{\rm mask}$, where $\omega$ is the per-galaxy weight and the sum is over all galaxies in the pixel. The galaxy density contrast is then given by $\hat{\delta}_{\rm g}(\hat{\mathbf{r}})=\hat{n}_{\rm gal}(\hat{\mathbf{r}})/\langle \hat{n}_{\rm gal} \rangle -1$, where $\hat{\mathbf{r}}$ denotes the sky position of the pixel centre 
and $\langle \hat{n}_{\rm gal} \rangle$ is the average effective number of galaxies per pixel over the survey footprint.
Despite applying the corrections with $\omega$ and $f_{\rm mask}$, we find that the galaxy density maps still have a large density gradient near the plane of our galaxy for all tomographic bins. We therefore apply an additional restrictive mask to exclude the region
$70^{\circ}<\textrm{RA}<330^{\circ}$
where this effect is most obvious. This cut reduces the fiducial DES Y3 footprint by about $20\%$. The number of galaxies is 267,021, 464,543, 709,092 for bins 1 - 3 respectively.

The \texttt{redMaGiC} catalogue was chosen for its photo-$z$ precision. However, there is a possible internal inconsistency in the \texttt{redMaGiC} catalogue: the ratio $X_{\rm lens}$ between the galaxy bias obtained from galaxy clustering and that obtained from the galaxy -- galaxy lensing two-point correlation functions is approximately $0.88$ \citep{2022PhRvD.106d3520P}.
This discrepancy results in a significant shift to the $S_8 (\equiv (\Omega_{\rm m}/0.3)^{0.5} \sigma_8)$ constraint compared to the DES baseline results using a magnitude limited lens sample \citep[\texttt{MagLim}; ][]{2021PhRvD.103d3503P}. 
\citet{2022PhRvD.106d3520P} discussed in detail the systematic effects that could lead to this inconsistency, and found that relaxing the cut in $\chi_{\rm max}^2$ could somewhat reduce the difference. 
In this paper, we also discuss the possible impact of $X_{\rm lens}$ on the measurements of galaxy bias.

\subsection{DES Y3 mass map}
\label{sec: mass map}

The DES Y3 weak lensing mass map is reconstructed from the DES Y3 \textsc{Gold} photometric galaxy sample \citep{2021ApJS..254...24S} and its shear catalogue \citep{2021MNRAS.504.4312G}. The source galaxy catalogue consists of 100,204,026 galaxies.
The \textsc{Gold} galaxies are split into four source tomographic bins\footnote{{The source binning version used is v0.4.}} with redshift distributions calibrated per bin using a Self-Organizing Map \citep{2019MNRAS.489..820B, 2021MNRAS.505.4249M}. 
The uncertainty in the source photometric redshift distribution is characterised via a set of discrete realisations using \textsc{HyperRank} \citep{2022MNRAS.511.2170C}. These samples are then used to propagate and marginalize over the possible photometric redshift errors.
The upper panel of Fig.~\ref{fig: Buzzard_redmagic_sim_nz} shows the average of the \textsc{HyperRank} realisations for the four source tomographic bins. Due to limited signal-to-noise ratio and the overlap with lens bins, we do not use the nearest two source bins.
For each source bin we create a HEALPix shear map in which each pixel has shear
\begin{equation}
    \gamma_{\rm obs}^{\nu}=\frac{\sum_{j}\epsilon^{\nu}_j w_j}{\bar{R}\sum_{j} w_j},
    \label{eq: shear map}
\end{equation}
where the sums are taken over all source galaxies in the pixel, $\epsilon^\nu$ is the shear field with component $\nu$, $w_j$ is a weight proportional to the inverse shear variance, and $\bar{R}$ is the average shear response from \textsc{Metacalibration} \citep{2017ApJ...841...24S, 2017arXiv170202600H}. 

The shear field $\gamma$ (of spin-weight 2) and the convergence field $\kappa$ (of spin-weight 0)
are related in spherical harmonic space via the Kaiser-Squires method \citep{1993ApJ...404..441K}:
\begin{equation}
    \hat{\gamma}_{\ell m}=-\sqrt{\frac{(\ell-1)(\ell+1)}{\ell(\ell+1)}}\hat{\kappa}_{\ell m}.
    \label{eq: KS}
\end{equation}
This allows $\kappa$ to be reconstructed from $\gamma$.
Let $\kappa_E$ and $\kappa_B$ be the real and imaginary parts, respectively, of $\kappa$; we refer to the E-mode, $\kappa_E$, as the reconstructed mass map, and hereafter drop the subscript $E$.
In the absence of systematic errors, $\gamma$ is curl-free and hence $\kappa_B=0$.

Within a Bayesian setting, \cite{2021MNRAS.505.4626J} reconstructed the DES Y3 mass maps using a likelihood derived from Eq.~\ref{eq: KS} and four different priors on $\kappa$: uniform prior (direct Kaiser-Squires inversion), Null B-mode prior, Gaussian prior (Wiener filter), and sparsity prior with the \textsc{Glimpse} algorithm \citep{2016A&A...591A...2L}. The different prior assumptions lead to reconstructed maps that are visually different, as shown in Fig. 10 of \cite{2021MNRAS.505.4626J}. 
We compared the stacked $\kappa$ profiles from these different reconstruction methods, concluding that for the scales used in this paper, these methods give consistent measurements. 
In this work, we adopt the Kaiser-Squires reconstruction for the mass maps.
This method is very susceptible to mask effects, which can induce a spurious B-mode \citep{2021MNRAS.505.4626J}. We account for this effect using forward-modeling with simulations.
The reconstructed map is also significantly degraded on small scales due to noise domination.
We subtract the mean field and smooth the maps with a Gaussian symmetric beam of angular scale $\sigma=20'$ to suppress the noise at small scales. 
{Throughout the analysis, we use the mass map with ${\rm nside}=512$ with the DES survey footprint mask.}

\begin{figure*}
     \centering
     \begin{subfigure}[b]{0.33\textwidth}
         \centering
         \includegraphics[width=\textwidth]{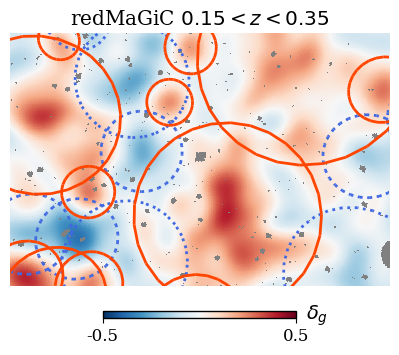}
     \end{subfigure}
     \hfill
     \begin{subfigure}[b]{0.33\textwidth}
         \centering
         \includegraphics[width=\textwidth]{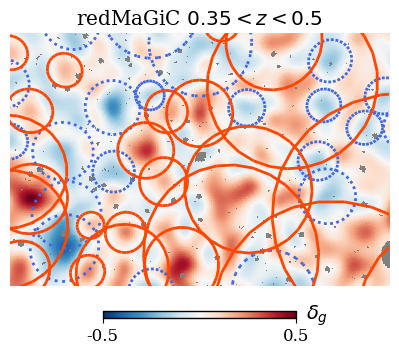}
     \end{subfigure}
     \hfill
     \begin{subfigure}[b]{0.33\textwidth}
         \centering
         \includegraphics[width=\textwidth]{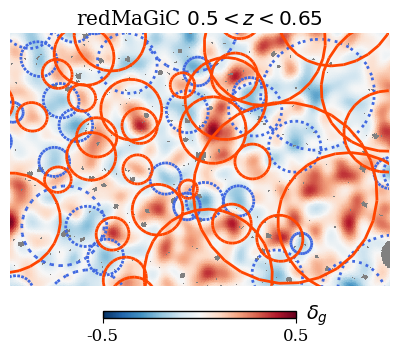}
     \end{subfigure}
        \caption{Cutouts of the DES Y3 \texttt{redMaGiC} galaxy density maps in three tomographic bins, centred on ${\rm RA}=40^{\circ}$, ${\rm Dec}=-35^{\circ}$, smoothed by a Gaussian filter with a fixed comoving scale $R_s=20\mpcoh$. Underdense regions are blue; overdense regions are red. The identified two-dimensional voids (resp. clusters) are shown by dashed (resp. solid) circles. The size of the cutout box is $50\times33 \,\text{deg}^2$ in Gnomonic projection. Notice that distortions are present at the edge of the boxes due to the large angular size.}
        \label{fig: DES_redmagic_y3_smthdens}
\end{figure*}

\begin{figure*}
    \centering
    \includegraphics[width=\textwidth]{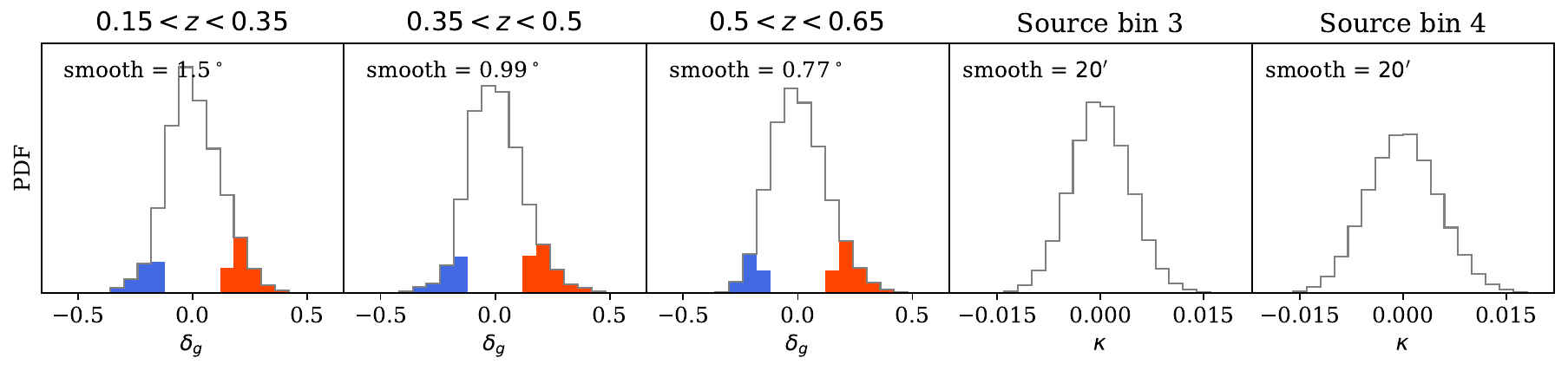}
    \caption{{The distribution of the smoothed galaxy density contrast for the void/cluster finding in the DES Y3 data (first three panels), and the smoothed kappa maps (last two panels). The smoothing angles are indicated in each panel for each lens/source redshift bin. The shaded regions are the pixels used for void (blue) and cluster (red) finding, defined by the threshold contrast $\delta_c$.}}
    \label{fig: delta_g_distribution}
\end{figure*}

\subsection{Gower Street simulations}
\label{sec: Dirac simulations}

The Gower Street simulations\footnote{Named for the street on which University College London is located.} \citep{2025MNRAS.536.1303J} are a suite of 791 full-sky $N$-body simulations generated using the \texttt{PKDGRAV3} code \citep{potter2017pkdgrav3}. Each simulation has a box of side length $1.25$ $h^{-1}{\rm Gpc}$, {containing $1080^3$ particles}, and a distinct $w$CDM cosmology varying seven cosmological parameters: $\Omega_{\rm m}, \sigma_8, w, \Omega_{\rm b}, H_0, n_s, m_{\nu}$. 
{The output of each simulation is 100 lightcone shells equally spaced in proper time between $z=49$ and $z=0$}, containing dark matter particle counts in HEALPix pixels with ${\rm nside}=2048$. Weak lensing shear is computed using ray-tracing with the Born approximation \citep{bornraytrace2020}. 
We select the samples with $w\geq-1.01$ and $0.2<\Omega_{\rm m}<0.5$.\footnote{We use simulations 1-128, 193-315, and 396-651 for our main analysis.} This gives a total of 415 simulations for forward modelling and computing the covariance matrix.
Finally, we take three additional simulations selected with the same criteria for validation tests {(these simulations have distinctive $\sigma_8$ values; see details in Sec.~\ref{sec: Testing the pipeline with the Gower Street Simulations})}.

We use the noisy shear maps, based on Gower Street simulations, constructed by \cite{2023arXiv231017557G}.
Shape noise in these maps is given by randomly rotating the ellipticity and weights of the DES Y3 shapes.
The shear multiplicative bias is drawn from a normal distribution, for each source tomographic bin, that is matched to the DES Y3 measurements.
Intrinsic alignment (IA) is implemented using the Non-linear Alignment Model \citep[NLA;][]{2004PhRvD..70f3526H, 2007NJPh....9..444B}, with the IA amplitude $a$ and redshift scaling $\eta$ drawn from uniform distributions: $a \in \mathcal{U}[-3, 3]$, $\eta \in \mathcal{U}[-5, 5]$. Photometric redshift uncertainties are included by drawing realizations of the source redshift distribution from \textsc{HyperRank}.
Source clustering is also included (see details in \cite{2023arXiv231017557G}).
Mass maps are reconstructed using the Kaiser-Squire method (as before), with the same Gaussian smoothing ($\sigma=20'$).

A key step in this analysis is to produce the mock \texttt{redMaGiC} galaxy density maps that match the number density, redshift distribution, and galaxy bias of the DES Y3 \texttt{redMaGiC} sample, using shells of dark matter density maps. This will be discussed in detail in Sec.~\ref{sec: Galaxy biases}.

\section{Methods}
\label{sec: methods}

\subsection{Trough and peak finding algorithm}
\label{sec: algo}

We use the two-dimensional void finding algorithm in \citet{2017MNRAS.465..746S} to find troughs and peaks of the projected galaxy density map, $\delta_{\rm g}$, for each lens tomographic bin.
{The algorithm operates in spherical geometry on HEALPix maps.}
The steps in the algorithm are:
\begin{enumerate}
    \item Smooth the density contrast field with a Gaussian beam with a fixed comoving scale $R_s$. The corresponding angular scale is $\theta_s=R_s/\chi_{\rm cen}$, where $\chi_{\rm cen}$ is the comoving distance to the centre of the bin. Higher redshift bins are therefore smoothed with a smaller angular kernel.
    \item Define a threshold density contrast $\delta_c$. Pixels with $\delta_{\rm g}<\delta_c$ (resp. $\delta_{\rm g}>\delta_c$) are identified as potential trough (resp. peak) centres.
    \item Find the pixel $p$ with the largest value of $|\delta_{\rm g}|$. Find the smallest ring, centered on $p$ and of fixed width $\Delta R$ {corresponding to the pixel resolution of the map}, that has a mean density contrast of zero. 
    The resulting ring and its enclosed points are then classified as an identified structure with radius $R_V$.
    \item Exclude all potential centers that fall within any identified structure.
    \item Repeat steps (iii) and (iv) until all potential centres are exhausted.
    \item Remove all structures that have more than 30\% of their area masked.
\end{enumerate}

{We refer to the identified structures as `voids' and `clusters' to distinguish them from the more general reference of troughs and peaks in the galaxy density field. }
This is a slight abuse of terminology, as `clusters' typically refer to galaxy clusters as found by (for example) friends-of-friends algorithms; such clusters are smaller than cosmic voids.
By contrast in our case the algorithm simply picks out the most extreme regions on the galaxy density maps, and hence these voids and clusters have similar scales.
{The last step removes $6, 13, 13$ voids and $0, 2, 2$ clusters from bins 1, 2, 3, respectively.}

The algorithm has two free parameters, $R_s$ and $\delta_c$.
The choice of $R_s$ affects the scale of the density fields being probed, as well as the signal-to-noise ratio of the measured profiles \citep{2017MNRAS.465.4166K}. 
In this analysis, we wish to measure linear scale bias, hence we set $R_s=20\mpcoh$; a larger $R_s$ would result in too sparse a sample in the first lens bin.
{These correspond to angular scales $\theta_s = 1.50^{\circ}, 0.99^{\circ}, 0.77^{\circ}$ for lens bins 1, 2, 3, respectively.}
The choice of $\delta_c$ controls how extreme are the centres of the identified structures; we choose $\delta_c$ to correspond to the $10\%$ most overdense (underdense) pixels on the smoothed density map for cluster (void) finding.
Fig.~\ref{fig: DES_redmagic_y3_smthdens} shows, on cutouts of the smoothed DES Y3 \texttt{redMaGiC} galaxy density maps, the voids and clusters identified using this procedure. {Changing the value of $\delta_c$ will generally not affect the deepest voids (peakest clusters) in the catalogue, but will add or remove ones close to the thresholds.}
{The $\delta_c$ values for the smoothed maps are $-0.15, -0.15, -0.16$ for voids and $0.16, 0.15, 0.16$ for clusters in lens bins 1, 2, 3, respectively. The fact that these values are very similar and symmetric for voids and clusters is the result of a large $\theta_s$ fixed in comoving scale.}
{Fig.~\ref{fig: delta_g_distribution} shows the distribution of the smoothed $\delta_g$ and $\kappa$ maps in the DES Y3 data. The smoothing angles for the galaxy density contrast are used for void/cluster finding, corresponding to a comoving scale of $R_s=20h^{-1}$Mpc, while the smoothing scale for the reconstructed $\kappa$ map is fixed to $\theta_s=20'$. In the figure the pixels used for void (cluster) finding are shaded blue (red) for the $\delta_g$ distribution. We can see a skewness in the smoothed $\delta_g$ distribution, while the distribution for $\kappa$ is more Gaussian.}

This algorithm is to be contrasted to some other two-dimensional structure finding methods, such as the counts-in-cell methods in \cite{2016MNRAS.455.3367G}. 
{Even though the algorithm is designed to identify independent peaks and troughs, there could still be deep, small voids found in a larger, shallower void, and vice versa for clusters.}
Note that these two-dimensional structures are also different from those found using a three-dimensional finder algorithm, such as watershed \citep{2007MNRAS.380..551P} or Voronoi tessellation \citep[\texttt{ZOBOV,}][]{2008MNRAS.386.2101N}.

We now describe `stacking'. Our stacked quantities are a function of an angular distance from structure centres; such angular distances are binned and we use $\theta$ to denote one such bin.
Given a pixelised field $f(p)$ on the sky (which may further depend on comoving distance $\chi$), we define a \textit{stacked average value taken over rings around superstructure centres}:

\begin{equation}
\langle f \rangle_{\rm Ring}(\theta^i) = \Avg_p f(p),
\label{eq: ringavg}
\end{equation}
where the average is taken over all pixels $p$ whose angular distance from the centre of an identified void (resp. cluster) in lens bin $i$ falls in angular distance bin $\theta$. A pixel will be counted twice (or more) if it is simultaneously angular distance $\theta$ from two (or more) structure centres.
Here, $f(p)$ can be the mass map $\kappa(p)$, the matter density contrast $\delta_{\rm m} (p)$, or the galaxy density contrast $\delta_{\rm g} (p)$.
We append a superscript $v$ or $c$ to distinguish between profiles derived from voids and clusters when the context is otherwise ambiguous.
We choose $\theta$ to be five linear angular bins in the range $[0.07^{\circ},11.31^{\circ}]$ (lens tomographic bin 1), $[0.05^{\circ},7.36^{\circ}]$ (bin 2), or $[0.04^{\circ},5.74^{\circ}]$ (bin 3).
These choices lead to roughly the same transverse distances. The lower bounds roughly correspond to the smoothing scale of the stacked maps, and the upper bounds roughly correspond to the size of the stacked profiles.

\subsection{Galaxy biases in the Gower Street Simulations}
\label{sec: Galaxy biases}

\begin{table}
	\centering
	\caption{Various galaxy bias definitions used in this paper. Notably, the input linear bias in the simulation, $b_{\rm input}$, can differ from the other biases (which are all consistent with each other on sufficiently large smoothing scales) due to the artifact that pixels with $\delta_{\rm g}<-1$ are set to $-1$ when making galaxy maps. Fig.~\ref{fig:bias_fm_estimators_comp} shows that $b_{\rm prof}$ has the same physical meaning as  $b_{C_{\ell}}$ and $b_{\RMS}$. }
	\label{tab: diff bias}
	\begin{tabular}{|p{0.4in}|p{2.5in}|}
	    \hline
		Symbol & Definition \\
		\hline
        \hline
		$b_{\rm input}$ & The linear galaxy bias used as input in the Gower Street simulations to compute the expected galaxy density, i.e. following the linear bias recipe of $\langle \delta_{\rm g} \rangle = b_{\rm input}\delta_{\rm m}$, but setting pixels with $\langle \delta_{\rm g} \rangle<-1$ to $-1$ (hence, the bias is not linear in nature). \\
        \hline
        $b_{C_{\ell}}$ & The galaxy bias defined as the ratio of the angular power spectra between galaxy-matter and matter-matter. The ratio is weighted by a Gaussian beam window function with a smoothing scale of $\sigma=20'$, consistent with the smoothing of the DES Y3 mass map. Hence, $b_{C_{\ell}} = \sum_{\ell} (C^{\rm gm}_{\ell} / C^{\rm mm}_{\ell}) W_{\ell}$\\
        \hline
        $b_{\RMS}$ & The galaxy bias defined as the Root-Mean-Square (RMS) of the galaxy and dark matter map maps smoothed with the same kernel as $b_{C_{\ell}}$. The bias is given by $b_{\RMS} = {\RMS}(\delta_{\rm g}(p)) / {\RMS}(\delta_{\rm m}(p))$.\\
        \hline
        $b^{v,c}_{\rm prof}$ & Ratio of the stacked galaxy and dark matter density profiles averaged over the $\theta$ bins, i.e. $b^{v,c}_{\rm prof}=\Avg_{\theta}(\langle \delta^{v,c}_{\rm g}\rangle_{\rm Ring}  / \langle \delta^{v,c}_{\rm m}  \rangle_{\rm Ring} )$. The superscripts $v,c$ denote voids and clusters. In this paper, we measure this bias via weak lensing mass map.
        \\
        \hline
	\end{tabular}
\end{table}

\begin{figure}
    \centering
    \includegraphics[width=\linewidth]{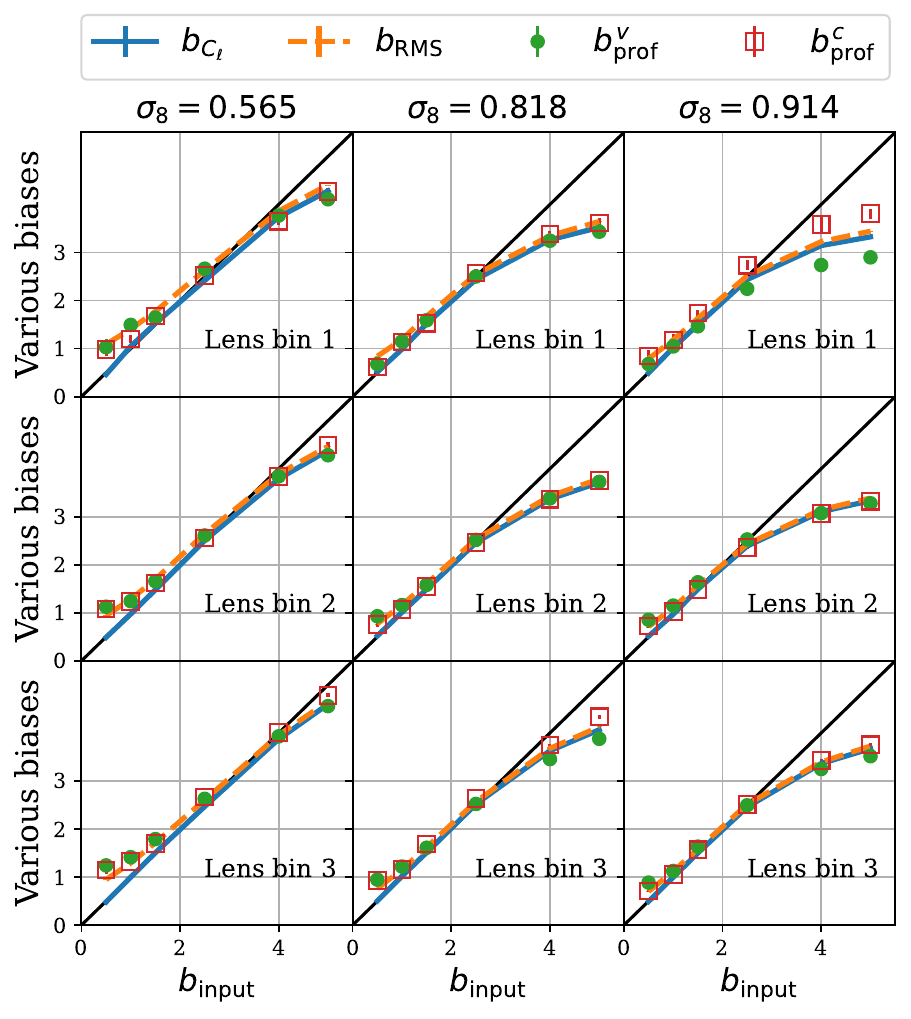}
    \caption{Various galaxy biases measured from the Gower Street simulations as a function of input bias in the simulation. The figures shows four sets of measured biases using the galaxy and dark matter density maps: $b_{C_{\ell}}$, the weighted ratio of angular power spectra (blue lines), $b_{\RMS}$, the ratio of the standard deviation of smoothed maps (orange dashed lines), and $b^{v,c}_{\rm MP}$, the profile ratio stacked around voids (green dots) and clusters (red open squares). The three columns show this relation for three simulations with increasing $\sigma_8$ values (left to right). The black line marks the diagonal. This provides a mapping between the input galaxy bias and the actual bias measured from the ratios of several quantities directly concerning the galaxy and matter field.
    The deviation from diagonal for small $b_{\rm input}$ is due to shot noise, and that for large $b_{\rm input}$ is due to the clipping of under-dense pixels.
    }
    \label{fig:bias_fm_estimators_comp}
\end{figure}


We describe the creation of mock \texttt{redMaGiC} galaxy maps from the Gower Street simulations.
A simulation gives the matter density contrast $\delta_{\rm m}(k, p)$ in pixelated shells; here $k$ is a shell (i.e. a redshift range) and $p$ is a pixel. 
We define the mean number of galaxies in a pixel and shell as $\lambda(k, p)= \bar{n}_{\rm gal}(k)\left[1+b_{\rm input} \ \delta_{\rm m}(k, p)\right]$ where $\bar{n}_{\rm gal}(k)$ is the average number of \texttt{redMaGiC} galaxies per pixel in redshift range $k$ in the actual DES Y3 data given a lens tomographic bin, and $b_{\rm input}$ is an input linear galaxy bias.
We sample the number of galaxies $N_{\rm gal}(p)$ from a Poisson distribution with mean $\sum_k \lambda(k, p)$ where the sum is over all shells $k$ in the redshift range of the lens tomographic bin. Finally, the galaxy density contrast field is given by $\delta_{\rm g}(p)= N_{\rm gal}(p)/\bar{N}_{\rm gal}-1$.

In pixels where $\sum_k \lambda(k, p) < 0$,
the Poisson mean becomes negative.
In keeping with common practice, 
we set these pixels to have zero galaxies i.e. to have a density contrast of $-1$.
However, this clipping operation alters the statistical properties of the generated biased field in a scale-dependent way,
and reduces the variance of the biased field. 
This is most problematic when $b_{\rm input}$ is large, or when the $\delta_{\rm m}$ field itself already has a large variance, e.g. dark matter maps at high resolution, at low redshifts, or with a high $\sigma_8$ value.
In these cases, the biased field may have a galaxy bias different to that of the simulation input bias $b_{\rm input}$, even in the linear regime. 

To investigate this effect, we also construct the corresponding `unbiased' dark matter density maps for each lens bin in the Gower Street simulations. This is given by $\delta_{\rm m}(p) = \sum_k \bar{n}_{\rm gal}(k) \delta_{\rm m}(k, p) / \sum_k \bar{n}_{\rm gal}(k)$, where as before the sum is over all shells $k$ in the redshift range of the lens tomographic bin.
We then measure the galaxy bias in the simulation using the $\delta_{\rm g}(p)$ and $\delta_{\rm m}(p)$ fields in the following ways (Table~\ref{tab: diff bias} provides a summary of these different bias definitions):

\begin{itemize}
    \item \textit{Angular power spectra ratio}:
    We measure galaxy bias from the matter and the (shot-noise subtracted) galaxy auto-correlations, $(C_{\ell}^{\rm gg}/C_{\ell}^{\rm mm}
    )^{1/2}$, and from the galaxy-matter cross-correlation, $C_{\ell}^{\rm gm}/C_{\ell}^{\rm mm}$. The auto- and cross-ratios are consistent, hence we only show cross-correlation results. Both the matter and galaxy density contrast maps are smoothed by a symmetric Gaussian beam $G_{\ell}(\sigma)$ with $\sigma=20'$, same as used for profile measurements, before measuring the angular power spectra $C_{\ell}$. This is to make sure the scales used are consistent between different bias measurements. We find that the ratios are mostly flat over a large $\ell$ range, between $\ell=[20,500]$, although for the highest $b_{\rm input}$ case, the ratio increases with $\ell$. {To get a single bias value from the measurement, we compute the average ratio between $\ell_{\rm min}=50$ and $\ell_{\rm max}=100,150,200$ for lens bins 1, 2, 3. The lower limit is to remove the effect of survey mask, and the upper limit roughly corresponds to $k_{\rm max}=0.1 h {\rm Mpc}^{-1}$, where the non-linearity of the density field becomes important. The uncertainty of the measurement is estimated as the standard deviation of the $\ell$-modes on the mean ratio within the range $\left(\ell_{\rm min}, \ell_{\rm max}\right)$.}
    \item \textit{RMS ratio}:
    We take the ratio of the Root-Mean-Square (RMS) of the pixel values of the smoothed galaxy and dark matter density maps, i.e. $b_{\RMS} = {\RMS}(\delta_{\rm g}(p)) / {\RMS}(\delta_{\rm m}(p))$ with the same smoothing kernel as above.
    {The uncertainty of the RMS is estimated via ${\rm RMS}/\sqrt{2(N_p-1)}$ (where $N_p$ is the number of pixels used to estimate the field RMS) and this is combined in quadrature to give the uncertainty on $b_{\rm RMS}$.}
    \item \textit{Profile ratio}:
    We measure the stacked dark matter density profiles, $\langle \delta_{\rm m}\rangle_{\rm Ring}(\theta)$, in addition to the galaxy density and $\kappa$ profiles, for the identified voids and clusters in the simulation using the algorithms of Sec.~\ref{sec: algo}. The bias is then measured by $b^{v,c}_{\rm MP} = \Avg_{\theta} ( \langle \delta_{\rm g}^{v,c}\rangle_{\rm Ring}  / \langle \delta_{\rm m}^{v,c}\rangle_{\rm Ring} )$, where the outer average is over the $\theta$ bins, and where the superscript $v,c$ stands for voids and clusters.
    {The uncertainty is estimated via Jackknife resampling by excluding one void (cluster) at a time when measuring the profiles.}
\end{itemize}


Fig.~\ref{fig:bias_fm_estimators_comp} shows, for three Gower Street simulations (with $\sigma_8=0.565, 0.818, 0.914$), how these various galaxy biases compare to the input bias $b_{\rm input}$ used in the simulations; six input biases in the range  $0.5 - 5.0$ were used.
As expected, the measured biases $b_{C_{\ell}}, b_{\RMS}, b^{v}_{\rm prof}, b^{c}_{\rm prof}$ are consistent with each other, but deviate from $b_{\rm input}$ significantly at high $b_{\rm input}$ values.
There, the measured biases `saturate' and increases little with $b_{\rm input}$. 
This trend is more significant at lower redshift bins, and for simulations with higher $\sigma_8$.
We note $b^v_{\rm prof}$ deviates from $b^c_{\rm prof}$ and becomes smaller than $b_{\RMS}$ and $b_{C_{\ell}}$ in bin 1 of the high $\sigma_8$ simulation. This could be because the most extreme underdensity values are capped, whereas those for the overdensity are not.
Below $b_{\rm input}\sim2$, $b_{C_{\ell}}$ agrees with $b_{\rm input}$ well, but other galaxy biases have higher values. This is because shot-noise of the galaxy field drops out in the $C_{\ell}^{\rm gm}$ measurements, but is still present in the other measurements. 
This shot-noise impact is larger at the lower bias end, partially because the signal-to-noise ratio is smaller, and partially due to the changes in the identification of clusters and voids: pixels close to but smaller than the threshold $\delta_c$ can be included in the finder algorithm.
Hence, it is important to differentiate the input, $b_{\rm input}$, from the measured galaxy biases on the biased fields in the simulations. 

\begin{table}
	\centering
	\caption{Number of structures identified on the \texttt{redMaGiC} galaxy density maps for the DES Y3 data and for one Gower Street simulation that had cosmology $\Omega_{\rm m}=0.290, \Omega_b=0.050, h=0.667, \sigma_8=0.766, w=-1.01$. The superscripts $v,c$ denotes voids and clusters, respectively.}
	\label{tab: superstructure number}
	\begin{tabular}{lccccc} 
	    \hline
		Lens bin & 1 & 2 & 3 \\
            & $0.1<z<0.35$ & $0.35<z<0.5$ & $0.5<z<0.65$ \\
		\hline
		$N^c$ (data) & 18 &  42&  76\\ 
		$N^v$ (data) & 13&  42&  63\\ 
            \hline
            $N^c$ (mock) & 16 & 50 & 81\\ 
            $N^v$ (mock) & 13 & 45 & 76\\ 
		\hline
	\end{tabular}
\end{table}

In our forward modelling approach, we choose $b_{\rm input}=1.7$ as the fiducial value for all 415 simulations. This particular value is chosen for two reasons: 1) it is close to the \texttt{redMaGiC} bias from the DES Y3 $3\times2$pt analysis \citep{2022PhRvD.105b3520A}, and 2) the agreement between various measured galaxy biases and $b_{\rm input}$ is reasonable, in that simulations with different $\sigma_8$ values have roughly the same measured bias.
In Sec.~\ref{sec: Testing the pipeline with the Gower Street Simulations}, we test the effect of different $b_{\rm input}$. 
We apply the void and cluster finding algorithm consistently on both the DES Y3 data and the Gower Street simulations. 
{Notice that the conversion from $R_s$ to angular scales in the finder algorithm assumes the simulation cosmology. However, in real data, this could distorted by the Alcock-Paczynski (AP) effect \citep{1979Natur.281..358A}, and the proper way to incorporate this is to convert using the same cosmology as the data in the forward modeling approach. We test the impact of the AP effect by re-analysing the actual data assuming the best-fit DES Y3 cosmology, instead of \textit{Planck}, finding that while there are slight changes to the superstructures found (due to different smoothing angles), the resultant profiles and their ratios are not significantly changed. Hence, we expect that our results will not be significantly impacted by the AP effect.}
Table~\ref{tab: superstructure number} shows the number of voids and clusters found in the DES Y3 data and in one Gower Street simulation that had reasonable cosmological parameters ($\Omega_{\rm m}=0.290, \Omega_b=0.050, h=0.667, \sigma_8=0.766, w=-1.01$).
The differences between the object numbers are consistent with Poisson noise, indicating that this fiducial choice for $b_{\rm input}$ is reasonable.

\subsection{Measurements of the profiles}

\begin{figure*}
     \begin{subfigure}[b]{\textwidth}
         \centering
         \includegraphics[width=\textwidth]{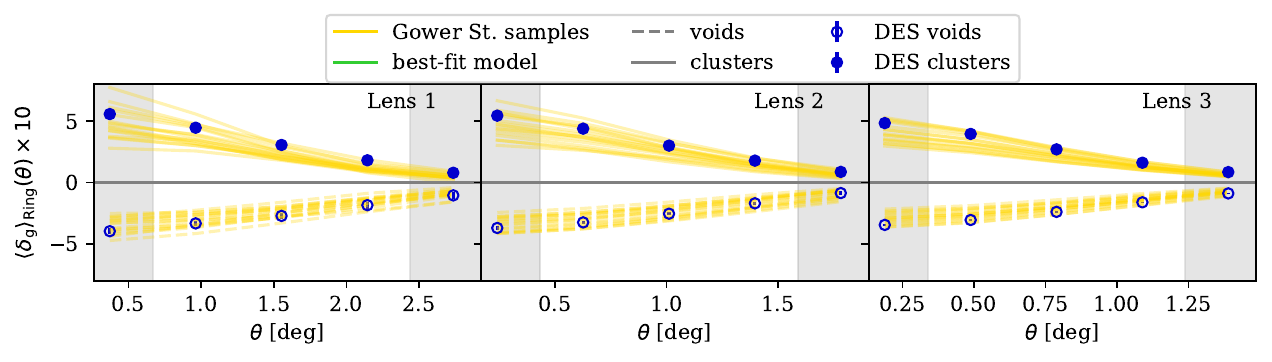}
     \end{subfigure}
    \hfill
    \begin{subfigure}[b]{\textwidth}
         \centering
         \includegraphics[width=\textwidth]{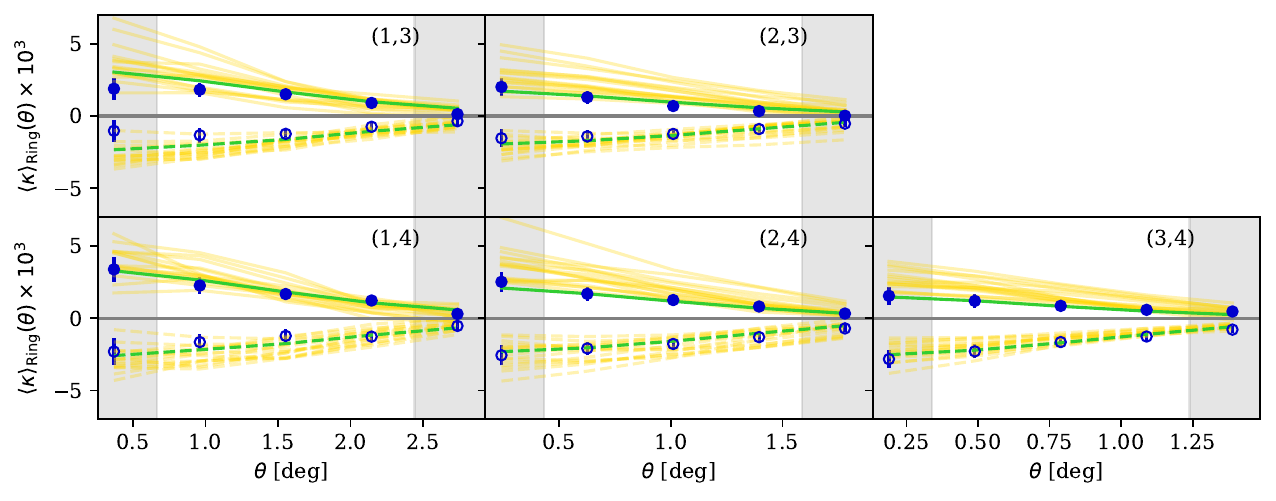}
     \end{subfigure}
        \caption{The stacked galaxy density profiles $\langle \delta_{\rm g} \rangle_{\rm Ring}(\theta)$ (upper panel) and the lensing profiles $\langle \kappa \rangle_{\rm Ring}(\theta)$ (lower panel) for voids (empty points and dashed lines) and clusters (filled points and solid lines), measured in the DES Y3 data (blue) and in 16 Gower Street simulations (yellow). The lens-source combination $(i,j)$ is indicated the upper right corner. The Gower Street samples have cosmological parameters within ranges $0.27\leq\Omega_{\rm m}\leq 0.32$ and $-1<w<-0.9$, and have an input linear bias of $b_{\rm input}=1.7$. For the data $\langle \kappa \rangle_{\rm Ring}(\theta)$ profiles, the solid and dashed green lines show the template model with the best-fit bias values.
        Grey shading denotes angular scales omitted from the final calculations.
        }
        \label{fig: results}
\end{figure*}

We measure the stacked profiles of voids and clusters of the galaxy density contrast and of the reconstructed mass map both for the DES Y3 data and for the Gower Street simulations. The galaxy density maps in all tomographic bins are smoothed with the same kernel as the mass map (a $\sigma=20'$ Gaussian beam) before stacking. We measure the profiles for all combinations of lens and source bin, but for the analysis use only five combinations that {have signal-to-noise ratios higher than ${\rm SNR}\sim3$ for the measurements at smallest $\theta$s} (namely $(i,j) = (1,3),(1,4),(2,3),(2,4),(3,4)$ for lens tomographic bin $i$ and source tomographic bin $j$).

Fig.~\ref{fig: results} shows these measurements both for the DES Y3 data and for 16 particular Gower Street simulations; these simulations have $0.27\leq\Omega_{\rm m }\leq 0.32$ and $-1<w<-0.9$, ranges that match \textit{Planck} \citep{2020A&A...641A...6P} and mean DES Y3 \citep{2022PhRvD.105b3520A} cosmologies. Only these two parameters are matched because the ratio between galaxy density and lensing profiles is most sensitive to their values (see Sec.~\ref{sec: cosmo}).
These realizations have a relatively wide $S_8$ range of $0.6<S_8<1.0$.
The error bars on the stacked $\kappa$ profile are taken from the diagonal of the covariance matrix (see Sec.~\ref{sec: Covariance from the Gower Street simulations}).
There is a reasonable agreement between the actual data and the simulation samples.
The stacked profiles of the clusters show a positive, decreasing amplitude from the centre (small $\theta$), dropping to nearly zero at a scale corresponding to a comoving size of $\sim 50 \mpcoh$. 
This is expected because these structures are found with an initial smoothing of $20\mpcoh$. Given their large comoving size, we shall also refer to these objects as `superstructures'.
The same feature, with a negative sign, can be seen in the stacked void profiles. 
In both data and simulations, the (absolute) amplitudes of the galaxy density profiles for voids are smaller than that for clusters, indicating a skewness in the galaxy density distribution even at these relatively large scales. A similar trend is present in the $\kappa$ profiles in Gower Street simulations. Interestingly, this is not the case for the DES Y3 data: the cluster profiles do not show a significantly higher amplitude than the void profiles.

\section{Modelling and analysis choices}
\label{sec: model}

\subsection{Weak lensing theory}

Given the three-dimensional galaxy number density $n(\mathbf{r})$ at position $\mathbf{r}$ within a volume $V$, one can define the corresponding galaxy density contrast, $\delta_{{\rm 3D,g}}(\mathbf{r})\equiv n(\mathbf{r})/\bar{n}-1$, where $\bar{n}$ is the average number density over the volume. The projected two-dimensional galaxy contrast is given by integrating the three-dimensional field along the line of sight, subject to the galaxy radial selection function, $p_{\rm lens}(\chi)$:
\begin{equation}
    \delta_{{\rm 2D,g}}(\hat{\mathbf{r}})=\int_0^{\chi_H} d\chi\, \delta_{{\rm 3D,g}}(\chi, \hat{\mathbf{r}})\, p_{\rm lens}(z) \frac{dz}{d\chi},
    \label{eq: density}
\end{equation}
where the three-dimensional vector $\mathbf{r}$ is now split into a radial comoving distance component $\chi$ and a two-dimensional angular component $\hat{\mathbf{r}}$. Here, $\chi_H$ is the comoving horizon scale and $p_{\rm lens}(z)$ is the normalised lens redshift distribution as defined before.

Under the Born approximation, the lensing convergence $\kappa$ induced by the foreground matter density field is
\begin{equation}
\begin{aligned}
    \kappa(\hat{\mathbf{r}}) = & \int_0^{\chi_H} d\chi\, \delta_{{\rm 3D,m}}(\chi, \hat{\mathbf{r}})  \frac{3H_0^2\Omega_{\rm m}}{2c^2}\frac{f_K(\chi)}{a(\chi)}\\
    & \times \int_{\chi}^{\chi_H}d\chi' p_{\rm source}(\chi')\frac{f_K(\chi'-\chi)}{f_K(\chi')},
    \label{eq: lensing}
\end{aligned}
\end{equation}
with $\delta_{{\rm 3D,m}}(\chi, \hat{\mathbf{r}})$ the three-dimensional dark matter density contrast, $H_0$ the Hubble constant, $\Omega_{\rm m}$ the matter fraction at $z=0$, $c$ the speed of light, $a(\chi)$ the expansion factor at $\chi$, $p_{\rm source}(\chi)$ the normalized radial source galaxy distribution, and $f_K(\chi)$ the transverse distance measure. The functional form of the latter depends on the curvature $\Omega_K$ and is $f_K(\chi)=\chi$ in a flat universe ($\Omega_K=0$).

Eqs.~\ref{eq: density} and \ref{eq: lensing} are connected through the galaxy and dark matter density contrast fields.
Under linear assumptions, the ratio of these fields is the galaxy bias i.e. $ \delta_{{\rm 3D,g}}(\mathbf{x}) =b \ \delta_{{\rm 3D,m}}(\mathbf{x})$.
The bias can in principle be a function of $\chi$ if it evolves with redshift; however, \texttt{redMaGiC} galaxies do not have a strong redshift evolution \citep{2022PhRvD.106d3520P} and so we assume a constant bias $b_i$ within lens tomographic bin $i$.


Combining Eq.~\ref{eq: ringavg}
with Eqs.~\ref{eq: density} and \ref{eq: lensing}, interchanging the order of pixel-averaging and $\chi$ integration, and applying the bias equation, shows 
$\langle \delta_{\rm g} \rangle^i_{\rm Ring}(\theta)$ and $\langle \kappa_j\rangle^{i}_{\rm Ring}(\theta)$ 
to have similar forms. Given a lens bin $i$ and the mass map for source bin $j$, we have
\begin{equation}
\label{eq: dens2}
\langle \delta_{\rm g} \rangle^i_{\rm Ring}(\theta)  =  b_i \ \int_0^{\chi_H} d\chi \ \langle \delta_{\rm 3D, m} \rangle^i_{\rm Ring}(\theta, \chi) \times p^i_{{\rm lens}}(z) \frac{dz}{d\chi},
\end{equation}
and
\begin{equation}
\label{eq: lensing2}
\langle \kappa_j\rangle^{i}_{\rm Ring}(\theta) = \int_0^{\chi_H} d\chi \ \langle \delta_{\rm 3D, m} \rangle^i_{\rm Ring}(\theta, \chi) \ \Phi_{j}(\chi) \ .
\end{equation}
Here
\begin{equation}
\Phi_{j}(\chi) = \frac{3H_0^2\Omega_{\rm m}}{2c^2} \frac{f_K(\chi)}{a(\chi)} \int_{\chi}^{\chi_H}d\chi' p^j_{{\rm source}}(\chi')\frac{f_K(\chi'-\chi)}{f_K(\chi')},
\end{equation}
and $\langle \delta_{\rm 3D, m} \rangle^i_{\rm Ring}(\theta, \chi)$ denotes taking the ring average of $\delta_{\rm 3D, m}$ at every $\chi$.


\begin{figure*}
     \centering
     \includegraphics[width=\textwidth]{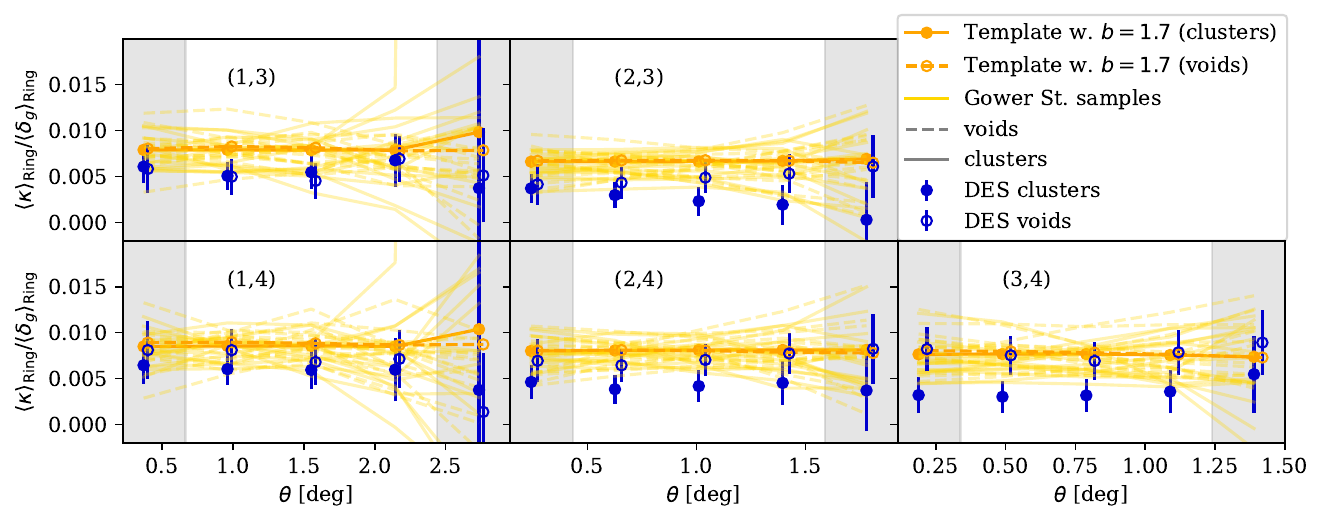}
    \caption{Similar to Fig.~\ref{fig: results}, but showing the measurements of the ratio between the stacked $\kappa$ profiles and the galaxy density contrast profiles, $\langle \kappa \rangle_{\rm Ring}(\theta)/\langle \delta_{\rm g}\rangle_{\rm Ring} (\theta)$, for the lens - source combination $(i,j)$.
    Values for voids (dashed lines) and clusters (solid lines) are slightly offset in $\theta$ for visual clarity.
    The measurements from the DES data are shown in blue, and the error bars are derived from the sample variance of the Gower Street simulations.
    Orange lines show the templates $F_{ij}$ constructed from the Gower Street simulations scaled to the \textit{Planck} cosmology, then divided by the input bias $1.7$ to roughly match to the measurements.
    Grey shading denotes angular scales omitted from the final calculations.
    The flatness of the measurements shows the two profiles to have similar $\theta$ dependence, while the vertical offset of the curves is attributed to galaxy bias and to differences in cosmology.
    } 
    \label{fig: F measurements}
\end{figure*}

\subsection{Template model}

Motivated by Eqs.~\ref{eq: dens2} and \ref{eq: lensing2} we write the ratio between the stacked mass map profile and the stacked galaxy density profile:
\begin{equation}
    \frac{\langle \kappa_j\rangle^i_{\rm Ring} (\theta)}{\langle \delta_{\rm g}\rangle^i_{\rm Ring}(\theta)} = \frac{F_{ij}(\cosmoparams,\theta)}{b_i},
    \label{eq: F}
\end{equation}
where $b_i$ is the galaxy bias (for lens bin $i$)
and $F_{ij}(\cosmoparams, \theta)$ is a `template' function (for lens bin $i$ and source bin $j$) that captures how the ratio depends on cosmological parameters $\cosmoparams$ and on $\theta$.

In theory $F$ will also depend on the exact shape of the stacked three-dimensional matter density contrast $\langle \delta_{\rm 3D, m} \rangle_{\rm Ring}(\theta, \chi)$.
However, in Appendix~\ref{sec: Model with the Buzzard mock} we show evidence that, to a good approximation, this quantity splits into the product of $p_{\rm lens}(z) dz/d\chi$ times a factor depending only on $\theta$. Were this approximation to be exact, then $\langle \delta_{\rm g} \rangle_{\rm Ring}(\theta)$ and $\langle \kappa \rangle_{\rm Ring}(\theta)$ would have the same sensitivity to $\theta$, and hence $F$ would have no $\theta$ sensitivity.
In practice, as discussed below, $F$ has mild $\theta$ sensitivity.
We do not model any further sensitivity of $F$ to the shape of $\langle \delta_{\rm 3D, m} \rangle_{\rm Ring}(\theta, \chi)$.

The actual $F$ can be systematically shifted from the theoretical prediction due to the reconstruction procedure. For example, the Kaiser-Squire method degrades the $\kappa$ signal on small scales, which can then make $F$ depend on scale. Systematic effects such as magnification, intrinsic alignment, and photometric redshift uncertainties can also impact the amplitude of $F$. This is explored further in Appendix~\ref{sec: systematics} using the toy analytical model from Appendix~\ref{sec: Model with the Buzzard mock}.


Fig.~\ref{fig: F measurements} shows the measurements of the ratio between the lensing and density profiles, $\langle \kappa\rangle_{\rm Ring} (\theta)/\langle \delta_{\rm g}\rangle_{\rm Ring}(\theta)$, for the DES Y3 data and for 16 Gower Street samples (the same samples as were used in Fig.~\ref{fig: results}).
Note that this ratio equals $b^{-1} F(\cosmoparams, \theta)$ within our model.
Over the angular scales of interest, the ratio of the two profiles has only a mild scale dependence (from which we infer that $F$ likewise has only mild scale dependence).
The increased scatter at larger angular scales results from $\langle \kappa\rangle_{\rm Ring} (\theta)$ and $\langle \delta_{\rm g}\rangle_{\rm Ring} (\theta)$ both being close to zero. 
In general, there is reasonable agreement between the simulations and the data. 
It is noticeable that most simulations have a larger amplitude of the ratio compared to the DES Y3 data, especially for clusters in lens bin 2 and 3. In light of Eq.~\ref{eq: F}, we see that this difference could be due either to a mismatch in cosmological parameters in $F$, or to a difference in the galaxy bias between the data and the simulations.

\subsubsection{Cosmological dependence of the templates}
\label{sec: cosmo}

\begin{figure}
         \centering
         \includegraphics[width=0.47\textwidth]{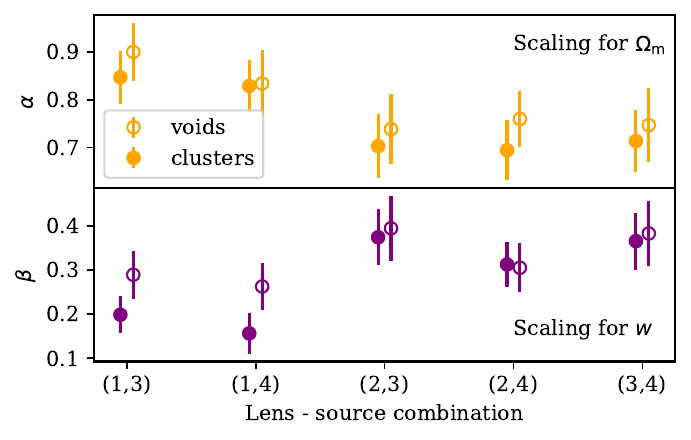}
        \caption{The cosmological dependence of $F_{ij}(\cosmoparams)$ on $\Omega_{\rm m}$ (upper panel) and $w$ (lower panel), assuming $F_{ij}(\Omega_{\rm m}, w)\propto \Omega_{\rm m}^{\alpha} w^{\beta}$. The values of $\alpha$ and $\beta$ for each lens-source combination for voids and clusters are measured from 312 Gower Street simulations, and the error bars are obtained from jackknife sampling. The lens-source combination is shown on the horizontal axis as $(i,j)$ for lens in bin $i$ and source in bin $j$.
        }
       \label{fig: scale}
\end{figure}

Taking a forward model approach, we compute a template $\tilde{F}_{ij}(\cosmoparams,\theta)$ from the Gower Street simulations.
To account for $b_i$ being not exactly equal to $b_{\rm input}=1.7$, we construct the template via
\begin{equation}
    \tilde{F}_{ij}(\cosmoparams,\theta) = \Avg_{\rm sim} s(\cosmoparams)\frac{\langle \kappa_j\rangle^i_{\rm Ring} (\theta)}{\langle \delta_{\rm m}\rangle^i_{\rm Ring}(\theta)}.
    \label{eq: F temp}
\end{equation}
The numerator is given by the noisy $\kappa$ maps including all the effects mentioned above. The denominator is given by the corresponding matter profiles. The average is taken over Gower Street simulations. The factor $s(\cosmoparams)$ allows scaling with cosmological parameters, such that $\tilde{F}_{ij}(\cosmoparams,\theta)$ is computed at a fixed cosmology. We describe the cosmological dependence below.

The limited signal-to-noise ratio of the profile measurements prevents us from simultaneously constraining both cosmology and galaxy bias; instead, we seek to measure galaxy bias at some fixed cosmology.
Therefore, when constructing the templates we need to scale out the cosmological dependence of $F_{ij}(\cosmoparams,\theta)$ in the Gower Street simulations. To this end we split the 415 simulations into 312 `training' samples, used to derive the cosmological dependence of $F_{ij}$, and 103 `testing' samples, used to validate this relation. 

We first measure the sensitivity of the `average over angular bins' $\Avg_{\theta} F_{ij}(\cosmoparams,\theta)$ to each of the cosmological parameters used in the Gower Street simulation suite.
We find that $F_{ij}$ is sensitive primarily to $\Omega_{\rm m}$ and to $w$, with little dependence on the other parameters. This is as expected: a) $\Omega_{\rm m}$ enters explicitly in Eq.~\ref{eq: lensing}; b) there is an implicit dependence on $\Omega_{\rm m}$ and $w$ through the distance -- redshift relation $a(\chi)$; c) perhaps not surprisingly, the $\sigma_8$ dependence cancels out in the ratio of profiles; this point was also discussed in \cite{2016MNRAS.462...35P, 2016MNRAS.459.3203C}. 

Given these observations, we assume a power-law scaling relation
\begin{equation}
    F_{ij}(\Omega_{\rm m}, w,\theta) =  F_{ij}(\Omega_{\rm m,0}, w_0,\theta)\left(\frac{\Omega_{\rm m}}{\Omega_{\rm m,0}}\right)^{\alpha}  \left(\frac{w}{w_0}\right)^{\beta}
    \label{eq: F scale}
\end{equation}
for some reference cosmology $(\Omega_{\rm m,0}, w_0)$\footnote{In practice, we typically convert between simulation cosmology and the {\it{Planck}} cosmology.}. We use least-squares to fit the parameters $\alpha$, $\beta$ and we estimate the uncertainty using jackknife resampling. 
Fig.~\ref{fig: scale} shows the measurements of the scaling parameters. 
The power for $\Omega_{\rm m}$ is around $\alpha=0.8$ for the first lens bin, and drops to around 0.7 for the other lens - source combinations.
The value agrees well between voids and clusters. The scaling parameter for $w$ is weaker: $\beta\sim0.3$.
Interestingly, combinations $(1,3)$ and $(1,4)$ hint at a slightly enhanced sensitivity to $w$ for voids compared to clusters. This can be expected because voids are dominated by dark energy.

We construct templates  $\tilde{F}_{ij}(\theta)$ at two fiducial $\Lambda$CDM (i.e. $w=-1$) cosmologies: the {\it Planck} best-fit cosmology \citep{2020A&A...641A...6P} with $\Omega_{\rm m}=0.315$, and the mean DES Y3 cosmology \citep{2022PhRvD.105b3520A} with $\Omega_{\rm m}=0.339$. The templates are computed by scaling the $F_{ij}$ of each simulation to these fiducial cosmologies using Eq.~\ref{eq: F scale}, and taking the average of the 312 `training' simulations. 
In Appendix~\ref{apdx: Model validation with Gower Street simulations}, we validate the scaling relation by scaling the template from {\it Planck} to the cosmology of each simulation in the test sample to recover their $\kappa$ profiles, finding agreement within $5\%$.
Fig.~\ref{fig: F measurements} shows (in orange) the templates $\tilde{F}_{ij}(\theta)$ for the \textit{Planck} cosmology, for all lens-source combinations and for voids and clusters separately. To match the $y$-axis, they have been multiplied by $1/b_{\rm input}$, where $b_{\rm input}=1.7$.

\subsection{Likelihood and covariance for bias estimation}
\label{sec: Covariance from the Gower Street simulations}


We use Bayesian inference to compute the posterior density of the galaxy bias $\hat{b}_{\rm prof}$ (denoted as such to distinguish it from other bias measurements mentioned in Table~\ref{tab: diff bias}).
By Bayes' theorem the posterior is given by 
\begin{equation}
    p(\hat{b}_{\rm prof}|\mathbf{d})\propto \mathcal{L}(\mathbf{d}|\hat{b}_{\rm prof}) \ \pi(\hat{b}_{\rm prof}),
\end{equation}
where $\mathbf{d}$ is the data, $\mathcal{L}(\mathbf{d}|\hat{b}_{\rm prof})$ is the likelihood, and $\pi(\hat{b}_{\rm prof})$ is the prior.

We adopt a uniform prior $\pi(\hat{b}_{\rm prof}) = \mathcal{U}[0.5,5]$ in each lens bin. {This prior follows that adopted in the DES Y3 two-point analysis; the upper limit is also set because the model is not validated for $b_{\rm prof}>5$.}

Our data $\mathbf{d}$ consists of the $\delta_{\rm g}$ and $\kappa$ profiles.
We consider five lens-source combinations: (1,3) and (1,4) for lens bin 1, (2,3) and (2,4) for lens bin 2, and (3,4) for lens bin 3. 
For all combinations, we use only three of our original five angular bins: 
the first bin is discarded because it is close to or below the smoothing scales, while the last bin is discarded due to the relatively large noise (and therefore unstable measurements of $F_{ij}(\cosmoparams,\theta)$ from the Gower Street simulations).
Thus, the length of the profile vectors is 6, 6, and 3 in lens bin 1, 2, and 3, respectively.

Motivated by Eq.~\ref{eq: F} we define $\mathbf{m}$, a quantity dependent on bias $\hat{b}_{\rm prof}$, data $\mathbf{d}$, and cosmological parameters $\cosmoparams$, to be:
\begin{equation}
    \mathbf{m}(\hat{b}_{\rm prof}, \mathbf{d}, \cosmoparams) \equiv \langle \kappa_j\rangle^i_{\rm Ring}(\theta) - \hat{b}_{\rm prof}^{-1}F_{ij}(\cosmoparams, \theta) \langle \delta_{\rm g} \rangle^i_{\rm Ring}(\theta).
\end{equation}
If $\mathbf{d}$ is Gaussian distributed then $\mathbf{m}$ will be as well (for fixed bias), and by Eq.~\ref{eq: F} it will have expected value zero.
We therefore assume a likelihood $\mathcal{L}(\mathbf{d}|\hat{b}_{\rm prof})\propto\exp(-\chi^2/2)$, where
\begin{equation}
    \chi^2= \mathbf{m}(\hat{b}_{\rm prof}, \mathbf{d}, \cosmoparams)^T \ \mathbfss{C}^{-1} \ \mathbf{m}(\hat{b}_{\rm prof}, \mathbf{d}, \cosmoparams)
\end{equation}
with $\mathbfss{C}$ the covariance matrix and where we have fixed $\cosmoparams$ to be either
the best-fit \textit{Planck} cosmology or the mean DES Y3 cosmology. 
We make the simplifying assumption of a fixed $\mathbfss{C}$ that is independent of $\hat{b}_{\rm prof}$. With this assumption we see that the likelihood, as a function of $\hat{b}_{\rm prof}^{-1}$ for fixed $\mathbf{d}$, will be Gaussian. Thus the likelihood -- and, given the flat prior, the posterior as well -- will be asymmetric in $\hat{b}_{\rm prof}$. We justify these assumptions, and demonstrate the natural appearance of $1/\hat{b}_{\rm prof}$ in Appendix~\ref{apdx: likelihood}.


The covariance matrix is constructed assuming that the noise does not depend on cosmology: 
\begin{equation}
    C_{pq} = \Avg_k [ m_p(b_{\rm input}, \mathbf{d}_k, \cosmoparams_k) \ m_q(b_{\rm input}, \mathbf{d}_k, \cosmoparams_k)],
    \label{eq: cov}
\end{equation}
where $k$ indexes the simulations and the average is taken over the 415 simulations. Subscripts $p$ and $q$ denote components of the $\mathbf{m}$ vector. For each simulation, $\mathbf{m}$ is evaluated using the data and cosmological parameters appropriate for this simulation, and using the fixed simulation bias $b_{\rm input}=1.7$. 
The covariance is only weakly dependent on the fitted bias, because the uncertainty is dominated by that of the $\kappa$ profile, rather than the galaxy density profile which scales with $1/\hat{b}_{\rm prof}$.

We also apply the correction to the inverse covariance to account for bias due to finite number of simulations \citep{2007A&A...464..399H}: $\mathbfss{C}^{-1} \rightarrow (N-P-2)/(N-1) \mathbfss{C}^{-1}$, where $P$ is the degrees of freedom (dof).
Appendix~\ref{apdx: covariance} shows the resulting covariance matrix for each lens bin.

\subsection{Validation with different $b_{\rm input}$ in Gower Street Simulations}
\label{sec: Testing the pipeline with the Gower Street Simulations}

\begin{figure*}
    \centering
    \includegraphics[width=\linewidth]{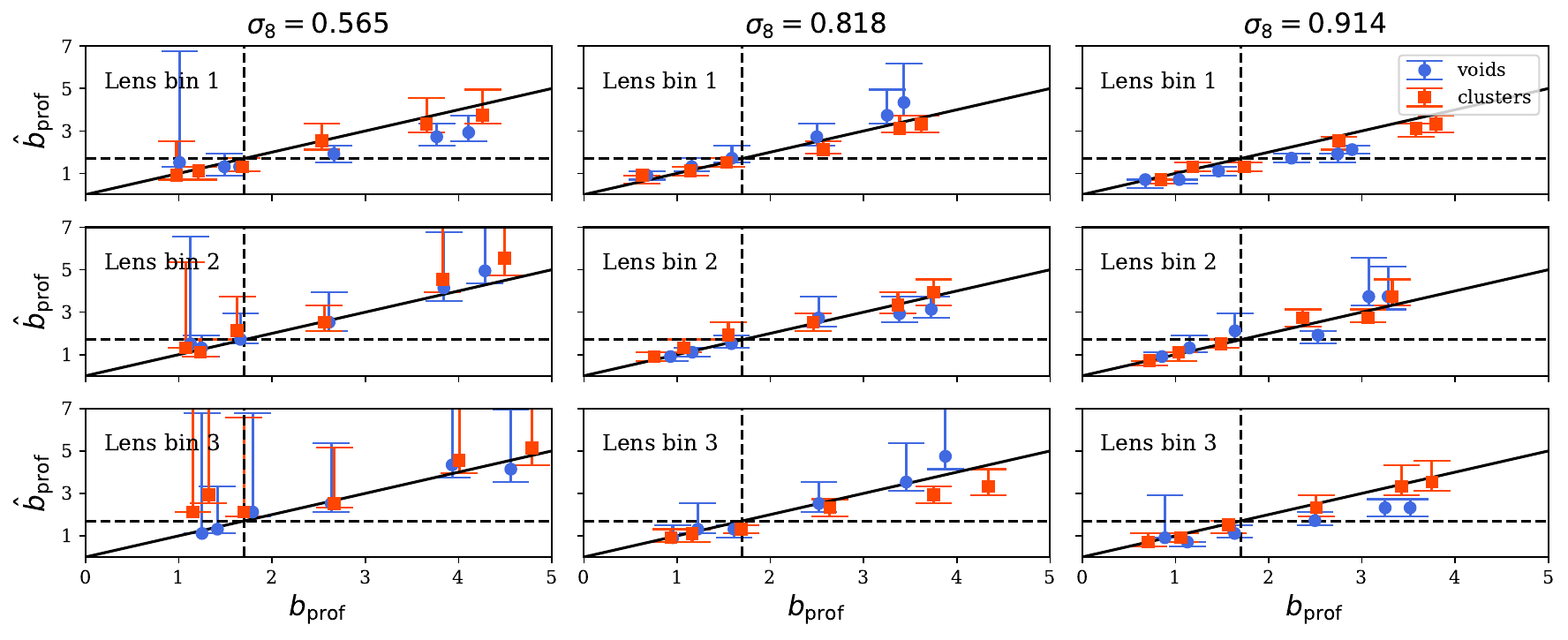}
    \caption{The estimated profile bias $\hat{b}_{\rm prof}$ from weak lensing ($y$-axis) versus the actual profile bias $b_{\rm prof}$ ($x$-axis), shown for three Gower Street simulations with varying $\sigma_8$ (columns).
    The various data points arise from using various $b_{\rm input}$ values (as shown in Fig.~\ref{fig:bias_fm_estimators_comp}).
    Voids are shown in blue and clusters in red. For $\hat{b}_{\rm prof}$, the maximum posterior value and the $68\%$ credible interval are shown for each point.
    The solid black diagonal line marks equality of the two biases, while the dashed black lines show $b=1.7$, from which the templates are computed.
    The two biases agree well over a wide range and for different $\sigma_8$ values.
    }
    \label{fig:bias_fm_estimators_comp-post}
\end{figure*}

The templates are constructed from simulations with $b_{\rm input}=1.7$. However, it might be that the template depends on the input bias of the simulations. For example, as discussed in Sec.~\ref{sec: Galaxy biases}, a low $b_{\rm input}$ can result in certain regions being misidentified as voids and clusters in the galaxy density map despite not corresponding to actual underdensities or overdensities in the dark matter map. This could reduce the amplitude and increase the uncertainty of the stacked $\kappa$ profile. 


In this section we perform a pipeline validation test by recovering the galaxy biases of Gower Street simulations with varying cosmology and input galaxy biases. We use the same three simulations as in Sec.~\ref{sec: Galaxy biases}, but with varying $b_{\rm input}$. Note that these simulations are not part of the 415 simulations used for computing the templates. To measure $\hat{b}_{\rm prof}$, we first scale the templates $\tilde{F}_{ij}(\cosmoparams,\theta)$ to the cosmology of each simulation according to Eq.~\ref{eq: F scale}, then use the measured galaxy density profile to compute the corresponding mass map profile, using Eq.~\ref{eq: F}.
The three simulations have the following cosmology, ordered from the lowest to the highest $\sigma_8$ values: $\Omega_{\rm m}=0.293, 0.313, 0.323$, $w=-0.841, -0.987, -0.630$, 
and $\sigma_8=0.565,0.818,0.914$.

To assess whether $\hat{b}_{\rm prof}$ is correctly recovered, we need to first clarify which bias in the simulation we are comparing to. As mentioned in Sec~\ref{sec: Galaxy biases}, the input galaxy bias differs from that measured from the ratio of angular power spectra, field RMS, or from the galaxy and dark matter profiles, especially at the high $b_{\rm input}$ end.
Given that the other bias measures agree reasonably well with each other, we choose to compare our $\hat{b}_{\rm prof}$ posterior with the bias measured from field RMS, $b_{\RMS}$, using the mapping as shown in Fig.~\ref{fig:bias_fm_estimators_comp}.


Fig.~\ref{fig:bias_fm_estimators_comp-post} shows the results of the validation test.
Each column refers to a particular Gower Street simulation, and each row refers to a particular lens tomographic bin.
The $y$-axis shows the best-fit $\hat{b}_{\rm prof}$ and the $68\%$ credible interval of its posterior distribution, and the $x$-axis shows the profile bias for voids (blue) and clusters (red) respectively, corresponding to different $b_{\rm input}$ for that simulation. 
In all cases, we are able to recover a $\hat{b}_{\rm prof}$ that is consistent with $b_{\rm prof}$, shown as the diagonal black lines in each panel.
There are some exceptions at the small end of the low-$\sigma_8$ simulation, and at the large end of the high-$\sigma_8$ simulation, where some $\hat{b}_{\rm prof}$ measurements statistically deviate from $b_{\rm prof}$.
This is particularly the case for the low $\sigma_8$ simulations.
The low-end deviation suggests that the low signal-to-noise ratio in this case causes deviations in the templates, and that the covariance matrix, which is obtained at the fiducial $b_{\rm input}$, may be underestimated. 
The high-end deviation, on the other hand, suggests that the covariance matrix may be slightly underestimated for these high $\sigma_8$ simulations, given that the signal-to-noise is higher than for lower $\sigma_8$ ones.
Nevertheless, despite this rather extreme case (in which $\sigma_8$ is far from the concordance cosmology and the bias is much different from typical \texttt{redMaGiC} galaxies), our pipeline is successful at recovering $\hat{b}_{\rm prof}$ over different cosmologies and over a wide range of $b_{\rm prof}$ in the simulations.


\section{Results}
\label{sec: Results}

\begin{figure*}
     \begin{subfigure}[b]{\textwidth}
         \centering
         \includegraphics[width=\textwidth]{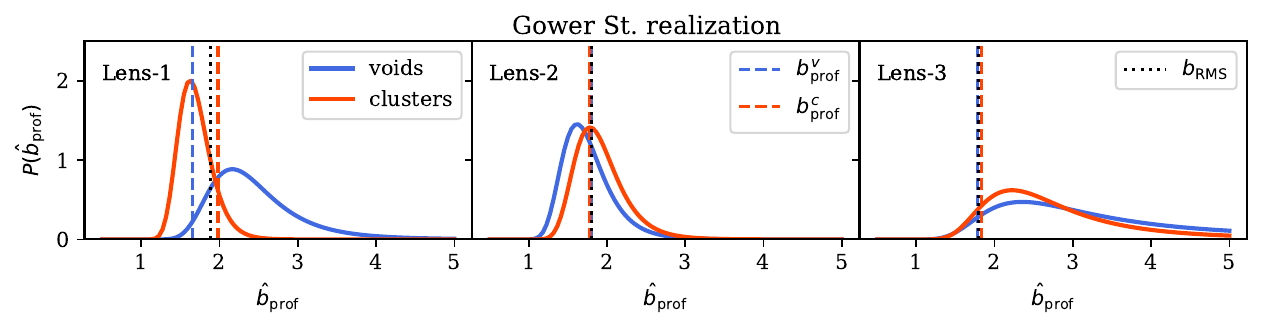}
     \end{subfigure}
    \hfill
    \begin{subfigure}[b]{\textwidth}
         \centering
         \includegraphics[width=\textwidth]{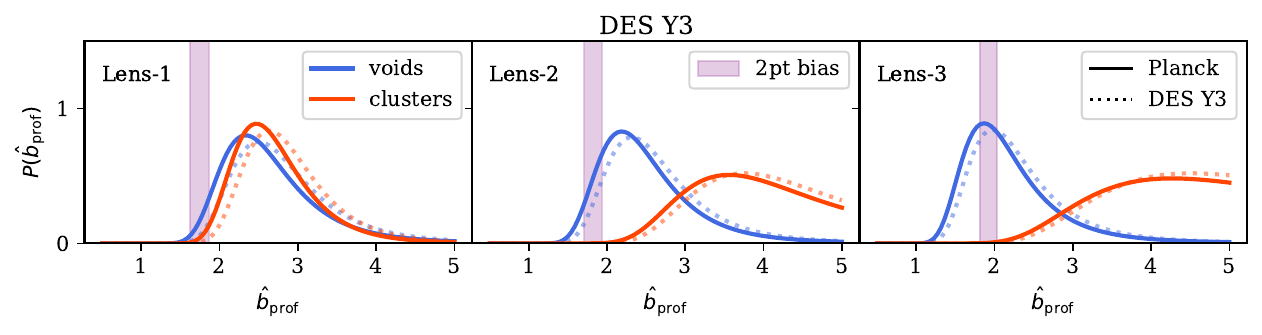}
     \end{subfigure}
        \caption{Posteriors for galaxy bias for voids (blue) and clusters (red), from a) one Gower Street simulation for which $\sigma_8=0.77$ (upper panel) and b) the DES Y3 data (lower panel). We adopt a uniform prior $\mathcal{U}[0.5,5]$ for the bias, even though the model constrains the reciprocal of the bias. Vertical dashed black lines in the upper panel mark $b_{\RMS}=1.84, 1.76, 1.76$, and the red and blue lines show $b^c_{\rm prof}=1.95,1.74,1.81$ and $b^v_{\rm prof}=1.62,1.79, 1.76$, respectively.
        The pink band in the lower panel indicates the $1\sigma$ credible interval of the bias determined from the DES Y3 $3 \times 2$pt analysis \citep{2022PhRvD.105b3520A}. 
        The posteriors were calculated assuming both \textit{Planck} cosmology (solid line) and DES Y3 best-fit cosmology (dashed line).
        }
        \label{fig: redMaGiC_model_deltag_kappa}
\end{figure*}

Fig.~\ref{fig: redMaGiC_model_deltag_kappa} shows the posterior probability densities for the galaxy biases measured (using the simulation-based template model) from a particular Gower Street simulation and from the actual DES Y3 data. 

The upper panel of Fig.~\ref{fig: redMaGiC_model_deltag_kappa} shows the results for a particular Gower Street simulation, one that has cosmological parameters $\Omega_{\rm m}=0.290, \Omega_b=0.050, h=0.667, \sigma_8=0.766, w=-1.01$ and that used $b_{\rm input}=1.7$. The RMS ratio biases are: $b_{\RMS} = 1.78, 1.76, 1.78$, and the profile ratios are $b^c_{\rm prof}=1.95,1.74,1.81$ and $b^v_{\rm prof}=1.62,1.79, 1.76$ for bins 1 -- 3, respectively.
The agreement of the measured profile bias with weak lensing is $1.94\sigma, 0.41\sigma, 1.82\sigma$ for voids, and $1.51\sigma, 0.01\sigma, 1.43\sigma$ for clusters, for lens bins 1 - 3 respectively.
The skewness of the posterior is due to the $1/\hat{b}_{\rm prof}$ factor from Eq.~\ref{eq: F}: the lower bound of the galaxy bias is relatively well constrained compared to the upper bound given the signal-to-noise ratio of the $\kappa$ profiles. 

The bottom panel of Fig.~\ref{fig: redMaGiC_model_deltag_kappa} shows the results for DES Y3 \texttt{redMaGiC} galaxies. The posteriors with the \textit{Planck} cosmology are shown by solid curves, with the maximum posterior values and 
$68\%$ credible intervals in lens bins 1 -- 3 measured to be 
\begin{equation}
    \begin{aligned}
    \hat{b}^v_{\rm prof}(\textit{Planck}) &= 2.32^{+0.86}_{-0.27}, 2.18^{+0.86}_{-0.23}, 1.86^{+0.82}_{-0.23} \quad (\text{voids}), \\
    \hat{b}^c_{\rm prof}(\textit{Planck}) &= 2.46^{+0.73}_{-0.27}, 3.55^{+0.96}_{-0.55}, 4.27^{+0.36}_{-1.14} \quad (\text{clusters}).
    \end{aligned}
\end{equation}

The $68\%$ credible intervals are determined by excluding the $16\%$ probability region from the tails of the distribution.
The posteriors assuming the mean DES Y3 cosmology are shown by dotted curves, with
\begin{equation}
    \begin{aligned}
        \hat{b}^v_{\rm prof} (\text{DES Y3}) & = 2.50^{+0.91}_{-0.27}, 2.32^{+0.86}_{-0.27}, 1.96^{+0.91}_{-0.23} \quad (\text{voids}), \\
        \hat{b}^c_{\rm prof}(\text{DES Y3}) & = 2.64^{+0.77}_{-0.27}, 3.77^{+0.77}_{-0.64}, 4.50_{-1.27}^{+0.18} \quad (\text{clusters}).
     \end{aligned}
\end{equation}

The maximum posterior of the DES cosmology is consistent with the \textit{Planck} posterior to $\sim0.1\sigma$.
The bias constraints assuming DES Y3 cosmology prefers a slightly larger value than \textit{Planck}, due to the larger $\Omega_{\rm m}$ values. 
Voids (blue) and clusters (red) have consistent bias measurements in the first bin, but the higher redshift bins prefer a slightly lower void bias and a significantly higher clusters bias. However, given that the posteriors are broad with a long tail towards large $\hat{b}_{\rm prof}$, the measurements are statistically consistent between density troughs and peaks within $\sim 2\sigma$.

The vertical shaded strips on the lower panel show the galaxy bias values with $1\sigma$ uncertainty from the DES Y3 $3\times2$pt combination \citep[Fig.~8 in][]{2022PhRvD.105b3520A}: $b_{3\times2{\rm pt}}=1.74\pm0.12, 1.82\pm0.11, 1.92\pm0.11$ for lens bin 1 -- 3, respectively. 
The consistency between our measured biases and the two-point bias assuming \textit{Planck} cosmology (DES Y3 cosmology) is $2.07\sigma$ ($2.58\sigma$), $1.36\sigma$ ($1.67\sigma$), $0.25\sigma$ ($0.56\sigma$) for voids and $2.78\sigma$ ($3.28\sigma$), $3.99\sigma$ ($4.39\sigma$), $3.28\sigma$ ($3.59\sigma$) for clusters, in lens bins 1 -- 3, respectively. 

It appears that most $\hat{b}_{\rm prof}$ measurements prefer larger bias values compared to that measured from two-point statistics, and for clusters at higher redshift bins the discrepancy is close to or greater than $3\sigma$. 

\cite{2022PhRvD.106d3520P} showed that the \texttt{redMaGiC} galaxy bias measured from galaxy-galaxy lensing is smaller compared to that from galaxy clustering. The discrepancy is quantified by the ratio of the two biases, 
\begin{equation}
    X_{\rm lens}\equiv b_{\rm ggl}/b_{\rm gc}=0.877^{+0.026}_{-0.019},
\end{equation}
where $ b_{\rm ggl}$ is fitted through the measured tangential shear $\gamma^{ij}_t(\theta)$ and theoretical matter correlation function $\xi^{ij}(\theta)_{\rm mm}$, via $\gamma^{ij}_t(\theta) = b_{\rm ggl}\, \xi^{ij}(\theta)_{\rm mm}$ in lens, source combination $(i,j)$. Similarly, $ b_{\rm gc}$ is measured from $w^{ii}(\theta) = b_{\rm gc}^2 \,\xi^{ii}(\theta)_{\rm mm}$, where $w^{ii}(\theta)$ is the lens galaxy angular correlation function. The cosmology of $\xi$ is fixed to the DES Y1 cosmology, which is very close to the Y3 cosmology. 
The measurement of $X_{\rm lens}\sim \langle \delta_g \kappa \rangle / \sqrt{\langle \delta_g \delta_g \rangle}$ resembles that of the template bias, where $1/\hat{b}_{\rm prof} \sim \langle \kappa \rangle / \langle \delta_g \rangle$. Hence, if there is a systematic effect that decorrelates the lensing signal from galaxy clustering, 
the measured $1/\hat{b}_{\rm prof}$ can be systematically smaller by the same factor of $X_{\rm lens}$. The $3\times2$pt bias value is consistent with a non-unity $X_{\rm lens}$ -- see e.g. the contour shifts from $2\times2$pt to $3\times2$pt statistics in Fig.~11 in \cite{2022PhRvD.105b3520A} -- i.e. the nominal bias is close to $b_{\rm gc}$.
If we account for the $X_{\rm lens}$ factor in the $\hat{b}_{\rm prof}$ posterior, the consistency becomes $1.36\sigma$ ($1.77\sigma$), $0.56\sigma$ ($0.86\sigma$), $0.25\sigma$ ($0.05\sigma$) for voids, and $1.77\sigma$ ($2.27\sigma$), $2.98\sigma$ ($3.38\sigma$), $2.58\sigma$ ($2.88\sigma$) for clusters, assuming \textit{Planck} cosmology (DES Y3 cosmology). The void biases are fully consistent, and the discrepancy in cluster bias in bins 2 and 3 is reduced (but is still significant).
{Notice that the $X_{\rm lens}$ is defined for the \textit{full} $\delta_{\rm g}$ field, whereas here we are identifying structures (peaks). If there is an additional fluctuation in the sample that is not correlated with LSS, this would manifest itself by giving more peaks, boosting the density (which will not be there in the kappa field). Hence the effect would also be sensitive to how $X_{\rm lens}$ scales with the density of the galaxies themselves.}

To test whether the residual difference in the measured biases is due to the choice of a particular cosmology, we repeat the measurement with varying $\Omega_m$ values (the primary dependence of the $\tilde{F}_{ij}(\cosmoparams, \theta)$ templates), and marginalize over this parameter.
We use a Gaussian prior centred at the best-fit DES Y3 value: $\Omega_{\rm m}\sim \mathcal{N}(0.339, 0.032)$. The resulting bias posteriors for the lens bins 1 -- 3 are
\begin{equation}
    \begin{aligned}
        & \hat{b}^v_{\rm prof} (\Omega_{\rm m} \,\text{marg.}) = 2.50_{-0.32}^{+0.96}, 2.32_{-0.27}^{+0.91}, 2.00_{-0.27}^{+0.86} \quad (\text{voids}),\\
        & \hat{b}^c_{\rm prof} (\Omega_{\rm m} \,\text{marg.}) = 2.64_{-0.32}^{+0.82}, 3.77_{-0.68}^{+0.77}, 4.50_{-1.32}^{+0.18} \quad (\text{clusters}),
    \end{aligned}
\end{equation}
without accounting for $X_{\rm lens}$. 
The error bars are slightly larger, but the overall conclusion is that marginalizing over $\Omega_{\rm m}$ does not change our results significantly.

{We do not consider baryonic feedback in this analysis. This effect refers to those baryonic feedback processes that redistribute the matter and hence affect the lensing results. This is more severe in overdense regions compared to underdense regions. An analysis of the baryonic effects on DES Y3 data by \cite{2022MNRAS.516.5355A} with a phenomenological model shows that the matter power spectrum can be affected up to $k\sim0.1 h {\rm Mpc}^{-1}$ at a few percent at $z=0$ and down to slightly smaller scales at $z=1$. For the lowest redshift bin, which potentially has the most impact, baryonic feedback starts to matter at a scale below $1^{\circ}$, which is smaller than the minimum scales of the analysis.}

\section{Conclusions}
\label{sec: conclusions}


We have measured galaxy bias at galaxy density troughs (voids) and peaks (clusters) in the DES Y3 \texttt{redMaGiC} galaxy catalogue, using their corresponding stacked profiles from the DES Y3 weak lensing mass maps as created using Kaiser-Squire reconstruction. 
The \texttt{redMaGiC} galaxies are split into three tomographic bins in $0.15<z<0.65$. A two-dimensional structure finder was applied on the galaxy density contrast maps smoothed at $20h^{-1}$Mpc to construct the void and cluster samples in each bin. The stacked, azimuthally averaged galaxy density profiles, $\langle \delta_{\rm g}  \rangle_{\rm Ring}(\theta)$, and lensing profiles, $\langle \kappa \rangle_{\rm Ring} (\theta)$, as a function of angular scales $\theta$ from the object centre, were measured for various lens-source combinations. 
The ratio of the two profiles is relatively flat as a function of $\theta$, indicating that in both overdense and underdense regions the {distribution of galaxies} and matter profiles have the same shape.
The amplitude of the ratio, on the other hand, depends mainly on the galaxy bias of the voids and clusters, with a mild dependence on cosmological parameters $\Omega_{\rm m}$ and $w$. 

To interpret these measurements, we take a simulation-based, forward-model approach with mocks constructed from 415 simulations from the Gower Street suite. To construct the mock galaxy catalogue, we sample galaxies with a Poisson distribution with linear bias on the underlying dark matter density maps. Pixels with negative mean galaxy counts are assigned zero galaxies, effectively setting a floor of $-1$ on $\delta_{\rm g}$.
We first quantify the difference (arising from this procedure) between this input linear bias, $b_{\rm input}$, and the galaxy bias on the $\delta_{\rm g}$ map (calculated various ways).
Then, for each combination of lens bin $i$ and source bin $j$, we construct a `template', $\tilde{F}_{ij}(\cosmoparams, \theta)$, and quantify its dependence on cosmological parameters. Realistic systematics and noise are included in the simulations, and are forward modelled to the templates. Finally, we scale the templates to two fiducial cosmologies: the best-fit \textit{Planck} 2018 cosmology and the mean DES Y3 cosmology, to fit the DES measurements. 

    
We measure the template-based bias, $\hat{b}_{\rm prof}$, in DES Y3 \texttt{redMaGiC} galaxies in the three lens bins, assuming \textit{Planck} cosmology, to be 
\begin{align*}
    \hat{b}^v_{\rm prof}(\textit{Planck}) &= 2.32^{+0.86}_{-0.27}, 2.18^{+0.86}_{-0.23}, 1.86^{+0.82}_{-0.23} \quad (\text{voids}), \\
    \hat{b}^c_{\rm prof}(\textit{Planck}) &= 2.46^{+0.73}_{-0.27}, 3.55^{+0.96}_{-0.55}, 4.27^{+0.36}_{-1.14} \quad (\text{clusters}).
\end{align*}

For DES Y3 cosmology, the measured biases are $3\% - 5\%$ larger.
In most cases, our measurements prefer a larger value compared to that in the DES Y3 $3\times2$pt analysis by $2\sigma - 4\sigma$, with the largest difference coming from clusters in bins 2 and 3. 
This value can be too large by a factor of $X_{\rm lens}\sim0.88$, as defined and measured in \cite{2022PhRvD.106d3520P}, given the potential decorrelation between \texttt{redMaGiC} galaxies and the lensing signal due to systematic effects. 
Upon accounting for this parameter, the difference with the cluster biases is reduced, and the void biases become fully consistent with the nominal $3\times2$pt value. We speculate that such decorrelation could be more severe at density peaks in higher redshift bins.
This issue will be revisited in future analyses, for example, DES Y6, LSST, and Euclid.

This paper presents a simple approach to measuring the correspondence of {distribution of galaxies} and matter at field level; there is certainly room for further exploration. 
On the simulation side, one can improve the linear biasing model such that the $b_{\rm input}$ corresponds to the bias given by e.g. angular power spectra. One possible avenue is using a power-law bias model, where $(1+\delta_{\rm g}) = (1+ \delta_{\rm m})^b$. While this scheme avoids the negative $n_{\rm gal}$ problem, it introduces very strong scale dependence of the biased field, even at linear scales. More investigation is thus needed on how to populate galaxies realistically on the projected dark matter shells for the Gower Street simulations.
On the measurement side, adopting a smaller smoothing scale for the superstructure finding algorithm would allow the probing of non-linear galaxy biases, which could then differ between density troughs and peaks and become scale-dependent.
One could further test whether the discrepancy with the DES nominal bias values are due to $X_{\rm lens}$ by repeating this analysis on the relaxed-$\chi^2_{\rm max}$ \texttt{redMaGiC} sample, or on the \texttt{MagLim} sample, where in both cases $X_{\rm lens}=1$.
Finally, instead of focusing on the $10\%$ most extreme structures, one could instead infer galaxy bias using the entire field. For example, this could be achieved via machine learning and simulation-based inference, as investigated by Williamson et al. (in prep.). These works will provide an interesting avenue for exploring field-level galaxy biases in forthcoming lensing data from e.g. DES Year 6, Rubin Observatory Legacy Survey of Space and Time (LSST), and the Euclid mission.
We leave these explorations for future work.

\section*{Author contributions} 

\textit{Data analysis and manuscript writing: } Q. Hang. \textit{Substantial suggestions on the project and the draft: } N. Jeffrey, L. Whiteway, O. Lahav. \textit{Generation of Gower Street simulations: } N. Jeffrey, L. Whiteway, M. Gatti. \textit{Generation of the Buzzard mock: } J. DeRose. \textit{Useful discussions: } J. Williamson, A. Kov\'{a}cs. The rest of the authors made the data products of DES Y3 available, including the Gold catalogue, the shear catalogue, redshifts, and simulations.

\section*{Acknowledegments}

QH and OL acknowledge STFC Consolidated Grant ST/R000476/1. OL also acknowledges visits to All Souls College and to the Physics Department, University of Oxford. NJ is supported by STFC Consolidated Grant ST/V000780/1 and ERC-selected UKRI Frontier Research Grant
EP/Y03015X/1. The Gower Street simulations were generated under the DiRAC project p153 ‘Likelihood-free inference with the Dark Energy Survey’
(ACSP255/ACSC1) using DiRAC (STFC) HPC facilities.
JW has been supported by the STFC UCL Centre for Doctoral Training in Data Intensive Science.
AK has been supported by a \emph{Lend\"ulet} excellence grant by the Hungarian Academy of Sciences (MTA), the European Union’s Horizon Europe research and innovation programme under the Marie Skłodowska-Curie grant agreement number 101130774, and the Hungarian Ministry of Innovation and Technology NRDI Office grant OTKA NN147550.

The authors would like to thank Christopher Davis, Elisa Legnani, and Shivam Pandey for serving as the DES internal review committee, whose helpful comments and suggestions improved the quality of this manuscript. The authors would also like to thank the anonymous referee in the MNRAS review process for constructive and insightful comments.

Funding for the DES Projects has been provided by the U.S. Department of Energy, the U.S. National Science Foundation, the Ministry of Science and Education of Spain, 
the Science and Technology Facilities Council of the United Kingdom, the Higher Education Funding Council for England, the National Center for Supercomputing 
Applications at the University of Illinois at Urbana-Champaign, the Kavli Institute of Cosmological Physics at the University of Chicago, 
the Center for Cosmology and Astro-Particle Physics at the Ohio State University,
the Mitchell Institute for Fundamental Physics and Astronomy at Texas A\&M University, Financiadora de Estudos e Projetos, 
Funda{\c c}{\~a}o Carlos Chagas Filho de Amparo {\`a} Pesquisa do Estado do Rio de Janeiro, Conselho Nacional de Desenvolvimento Cient{\'i}fico e Tecnol{\'o}gico and 
the Minist{\'e}rio da Ci{\^e}ncia, Tecnologia e Inova{\c c}{\~a}o, the Deutsche Forschungsgemeinschaft and the Collaborating Institutions in the Dark Energy Survey. 

The Collaborating Institutions are Argonne National Laboratory, the University of California at Santa Cruz, the University of Cambridge, Centro de Investigaciones Energ{\'e}ticas, 
Medioambientales y Tecnol{\'o}gicas-Madrid, the University of Chicago, University College London, the DES-Brazil Consortium, the University of Edinburgh, 
the Eidgen{\"o}ssische Technische Hochschule (ETH) Z{\"u}rich, 
Fermi National Accelerator Laboratory, the University of Illinois at Urbana-Champaign, the Institut de Ci{\`e}ncies de l'Espai (IEEC/CSIC), 
the Institut de F{\'i}sica d'Altes Energies, Lawrence Berkeley National Laboratory, the Ludwig-Maximilians Universit{\"a}t M{\"u}nchen and the associated Excellence Cluster Universe, 
the University of Michigan, NSF NOIRLab, the University of Nottingham, The Ohio State University, the University of Pennsylvania, the University of Portsmouth, 
SLAC National Accelerator Laboratory, Stanford University, the University of Sussex, Texas A\&M University, and the OzDES Membership Consortium.

Based in part on observations at NSF Cerro Tololo Inter-American Observatory at NSF NOIRLab (NOIRLab Prop. ID 2012B-0001; PI: J. Frieman), which is managed by the Association of Universities for Research in Astronomy (AURA) under a cooperative agreement with the National Science Foundation.

The DES data management system is supported by the National Science Foundation under Grant Numbers AST-1138766 and AST-1536171.
The DES participants from Spanish institutions are partially supported by MICINN under grants PID2021-123012, PID2021-128989 PID2022-141079, SEV-2016-0588, CEX2020-001058-M and CEX2020-001007-S, some of which include ERDF funds from the European Union. IFAE is partially funded by the CERCA program of the Generalitat de Catalunya.

We  acknowledge support from the Brazilian Instituto Nacional de Ci\^encia
e Tecnologia (INCT) do e-Universo (CNPq grant 465376/2014-2).

This document was prepared by the DES Collaboration using the resources of the Fermi National Accelerator Laboratory (Fermilab), a U.S. Department of Energy, Office of Science, Office of High Energy Physics HEP User Facility. Fermilab is managed by Fermi Forward Discovery Group, LLC, acting under Contract No. 89243024CSC000002.

\section*{Data Availability}

The code and data used for this paper are available upon reasonable request.



\bibliographystyle{mnras}
\bibliography{main} 


\section*{Affliations}

$^{1}$ Department of Physics \& Astronomy, University College London, Gower Street, London, WC1E 6BT, UK\\
$^{2}$ Kavli Institute for Cosmological Physics, University of Chicago, Chicago, IL 60637, USA\\
$^{3}$ Lawrence Berkeley National Laboratory, 1 Cyclotron Road, Berkeley, CA 94720, USA\\
$^{4}$ MTA--CSFK \emph{Lend\"ulet} ``Momentum'' Large-Scale Structure (LSS) Research Group, Konkoly Thege Mikl\'os \'ut 15-17, H-1121 Budapest, Hungary\\
$^{5}$ Konkoly Observatory, HUN-REN Research Centre for Astronomy and Earth Sciences, Konkoly Thege Mikl\'os {\'u}t 15-17, H-1121 Budapest, Hungary\\
$^{6}$ Institute of Space Sciences (ICE, CSIC),  Campus UAB, Carrer de Can Magrans, s/n,  08193 Barcelona, Spain\\
$^{7}$ Department of Astrophysical Sciences, Princeton University, Peyton Hall, Princeton, NJ 08544, USA\\
$^{8}$ Physics Department, 2320 Chamberlin Hall, University of Wisconsin-Madison, 1150 University Avenue Madison, WI  53706-1390\\
$^{9}$ Argonne National Laboratory, 9700 South Cass Avenue, Lemont, IL 60439, USA\\
$^{10}$ Department of Physics and Astronomy, University of Pennsylvania, Philadelphia, PA 19104, USA\\
$^{11}$ Department of Physics, Carnegie Mellon University, Pittsburgh, Pennsylvania 15312, USA\\
$^{12}$ NSF AI Planning Institute for Physics of the Future, Carnegie Mellon University, Pittsburgh, PA 15213, USA\\
$^{13}$ Instituto de Astrofisica de Canarias, E-38205 La Laguna, Tenerife, Spain\\
$^{14}$ Laborat\'orio Interinstitucional de e-Astronomia - LIneA, Av. Pastor Martin Luther King Jr, 126 Del Castilho, Nova Am\'erica Offices, Torre 3000/sala 817 CEP: 20765-000, Brazil\\
$^{15}$ Universidad de La Laguna, Dpto. Astrofísica, E-38206 La Laguna, Tenerife, Spain\\
$^{16}$ Center for Astrophysical Surveys, National Center for Supercomputing Applications, 1205 West Clark St., Urbana, IL 61801, USA\\
$^{17}$ Department of Astronomy, University of Illinois at Urbana-Champaign, 1002 W. Green Street, Urbana, IL 61801, USA\\
$^{18}$ Department of Astronomy and Astrophysics, University of Chicago, Chicago, IL 60637, USA\\
$^{19}$ Department of Physics, Duke University Durham, NC 27708, USA\\
$^{20}$ NASA Goddard Space Flight Center, 8800 Greenbelt Rd, Greenbelt, MD 20771, USA\\
$^{21}$ Fermi National Accelerator Laboratory, P. O. Box 500, Batavia, IL 60510, USA\\
$^{22}$ Universit\'e Grenoble Alpes, CNRS, LPSC-IN2P3, 38000 Grenoble, France\\
$^{23}$ Department of Physics and Astronomy, University of Waterloo, 200 University Ave W, Waterloo, ON N2L 3G1, Canada\\
$^{24}$ California Institute of Technology, 1200 East California Blvd, MC 249-17, Pasadena, CA 91125, USA\\
$^{25}$ SLAC National Accelerator Laboratory, Menlo Park, CA 94025, USA\\
$^{26}$ University Observatory, LMU Faculty of Physics, Scheinerstr. 1, 81679 Munich, Germany\\
$^{27}$ School of Physics and Astronomy, Cardiff University, CF24 3AA, UK\\
$^{28}$ Department of Applied Mathematics and Theoretical Physics, University of Cambridge, Cambridge CB3 0WA, UK\\
$^{29}$ Kavli Institute for Particle Astrophysics \& Cosmology, P. O. Box 2450, Stanford University, Stanford, CA 94305, USA\\
$^{30}$ Instituto de F\'isica Gleb Wataghin, Universidade Estadual de Campinas, 13083-859, Campinas, SP, Brazil\\
$^{31}$ Nordita, KTH Royal Institute of Technology and Stockholm University, Hannes Alfv\'ens v\"ag 12, SE-10691 Stockholm, Sweden\\
$^{32}$ Department of Physics, University of Genova and INFN, Via Dodecaneso 33, 16146, Genova, Italy\\
$^{33}$ Jodrell Bank Center for Astrophysics, School of Physics and Astronomy, University of Manchester, Oxford Road, Manchester, M13 9PL, UK\\
$^{34}$ Centro de Investigaciones Energ\'eticas, Medioambientales y Tecnol\'ogicas (CIEMAT), Madrid, Spain\\
$^{35}$ Brookhaven National Laboratory, Bldg 510, Upton, NY 11973, USA\\
$^{36}$ Department of Physics and Astronomy, Stony Brook University, Stony Brook, NY 11794, USA\\
$^{37}$ Institut de Recherche en Astrophysique et Plan\'etologie (IRAP), Universit\'e de Toulouse, CNRS, UPS, CNES, 14 Av. Edouard Belin, 31400 Toulouse, France\\
$^{38}$ Department of Physics, Stanford University, 382 Via Pueblo Mall, Stanford, CA 94305, USA\\
$^{39}$ INAF-Osservatorio Astronomico di Trieste, via G. B. Tiepolo 11, I-34143 Trieste, Italy\\
$^{40}$ Department of Physics, University of Michigan, Ann Arbor, MI 48109, USA\\
$^{41}$ Physik-Institut, University of Zürich, Winterthurerstrasse 190, CH-8057 Zürich, Switzerland\\
$^{42}$ Institute of Cosmology and Gravitation, University of Portsmouth, Portsmouth, PO1 3FX, UK\\
$^{43}$ Department of Physics, Northeastern University, Boston, MA 02115, USA\\
$^{44}$ Institut de F\'{\i}sica d'Altes Energies (IFAE), The Barcelona Institute of Science and Technology, Campus UAB, 08193 Bellaterra (Barcelona) Spain\\
$^{45}$ Physics Department, William Jewell College, Liberty, MO, 64068\\
$^{46}$ Institut d'Estudis Espacials de Catalunya (IEEC), 08034 Barcelona, Spain\\
$^{47}$ Hamburger Sternwarte, Universit\"{a}t Hamburg, Gojenbergsweg 112, 21029 Hamburg, Germany\\
$^{48}$ School of Mathematics and Physics, University of Queensland,  Brisbane, QLD 4072, Australia\\
$^{49}$ Department of Physics, IIT Hyderabad, Kandi, Telangana 502285, India\\
$^{50}$ Santa Cruz Institute for Particle Physics, Santa Cruz, CA 95064, USA\\
$^{51}$ Center for Cosmology and Astro-Particle Physics, The Ohio State University, Columbus, OH 43210, USA\\
$^{52}$ Department of Physics, The Ohio State University, Columbus, OH 43210, USA\\
$^{53}$ Australian Astronomical Optics, Macquarie University, North Ryde, NSW 2113, Australia\\
$^{54}$ Lowell Observatory, 1400 Mars Hill Rd, Flagstaff, AZ 86001, USA\\
$^{55}$ Jet Propulsion Laboratory, California Institute of Technology, 4800 Oak Grove Dr., Pasadena, CA 91109, USA\\
$^{56}$ George P. and Cynthia Woods Mitchell Institute for Fundamental Physics and Astronomy, and Department of Physics and Astronomy, Texas A\&M University, College Station, TX 77843,  USA\\
$^{57}$ Instituci\'o Catalana de Recerca i Estudis Avan\c{c}ats, E-08010 Barcelona, Spain\\
$^{58}$ Ruhr University Bochum, Faculty of Physics and Astronomy, Astronomical Institute, German Centre for Cosmological Lensing, 44780 Bochum, Germany\\
$^{59}$ Physics Department, Lancaster University, Lancaster, LA1 4YB, UK\\
$^{60}$ Computer Science and Mathematics Division, Oak Ridge National Laboratory, Oak Ridge, TN 37831\\


\appendix

\section{Analytic toy model with the \texttt{Buzzard} mock}

\label{sec: Model with the Buzzard mock}
\begin{figure*}
         \centering
         \includegraphics[width=\textwidth]{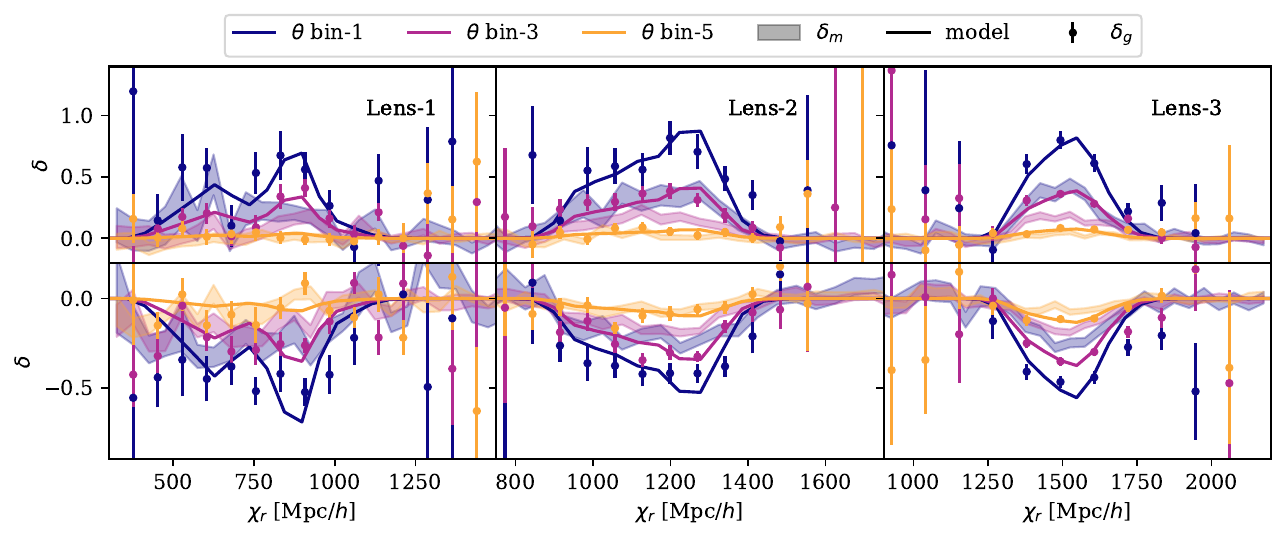}
        \caption{Stacked galaxy density contrast for voids (upper panel) and clusters (lower panel) measured on rings of angular scale $\theta$ in the three lens tomographic bins in the \texttt{Buzzard} simulation. The solid lines show the approximation in Eq.~\ref{eq: model} using the true redshift distribution for the lens sample and the measured galaxy two-dimensional density profiles. The shaded regions show the same for the underlying matter distribution. Error bars are given by jackknife sampling.}
        \label{fig: Buzzard_redmagic_sim_3D_stacked_0.35_z_0.5}
\end{figure*}

\begin{figure*}
	\includegraphics[width=\textwidth]{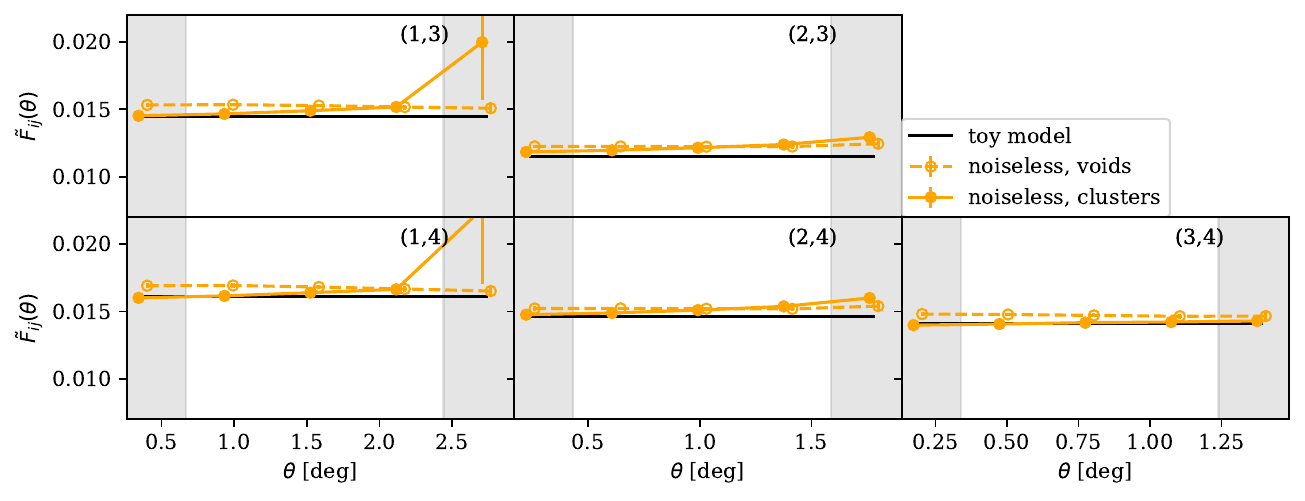}
    \caption{The noiseless template $\tilde{F}_{ij}(\theta)$ from the toy model prediction (black) and from the Gower Street simulations (yellow) at \textit{Planck} cosmology.
    The noiseless template from the simulation is constructed using the true $\kappa$ maps. Voids and clusters are denoted by dashed and solid lines respectively.}
    \label{fig: F_cosmo_buzzard}
\end{figure*}

As mentioned in Sec.~\ref{sec: model}, Eqs.~\ref{eq: dens2} and \ref{eq: lensing2} can be computed analytically, if one knows the line-of-sight structure of the stacked voids and clusters, i.e. $\langle \delta_{\rm 3D, g}\rangle_{\rm Ring} (\chi, \theta)$ and $\langle \delta_{\rm 3D, m}\rangle_{\rm Ring}(\chi, \theta)$.
In this section, we construct a toy model for the line-of-sight distribution of the stacked clusters and voids. To do so, we use the DES Y3 \texttt{Buzzard} mock \citep{2022PhRvD.105l3520D} that has matched galaxy selection to the DES Y3 \texttt{redMaGiC} and source samples. 
The \texttt{Buzzard} mock is an $N$-body simulation that assumes a flat $\Lambda$CDM cosmology with WMAP7 \citep{2011ApJS..192...18K} parameters: $\Omega_{\rm m}=0.286$, $\Omega_b=0.046$, $h=0.7$, $n_s=0.96$, and $\sigma_8=0.82$. It is chosen for this exercise because the galaxy sample is populated more realistically using the semi-analytic \texttt{Addgals} algorithm \citep{2022ApJ...931..145W}.
We use the mock \texttt{redMaGiC} catalogue with matched redshift distribution in each tomographic bin.
We additionally use the underlying dark matter number density in 80 shells with a comoving width of $50\mpcoh$. For each shell, we convert the number of dark matter particles $n_{\rm m}$ (HEALPix pixelized with ${\rm nside}=512$) to matter density contrast via $\delta_{\rm m}=n_{\rm m}/\bar{n}_{\rm m}-1$. Following the same algorithm as in Sec.~\ref{sec: algo}, we find void and cluster centres on the mock galaxy density maps.

We subdivide each lens tomographic bin into ten sub-bins of equal comoving distance. For each sub-bin $j$ in tomographic bin $I$ we compute the galaxy number density for a given pixel using $\delta_{{\rm g},j}^I=n^I_j/\bar{n}^I_j-1$, where $n^I_j$ is the number of galaxies in the sub-bin in that pixel, and $\bar{n}^I_j$ the averaged galaxy number over all pixels within the mask.
For each of the sub-bins, we measure the stacked cluster and void profiles as a function of angular scales $\theta$, on both the galaxy and dark matter density contrast maps, following the procedure described in Sec.~\ref{sec: methods}.
In this way, we `step through' the line-of-sight structure of $\langle \delta_{\rm 3D, g}\rangle_{\rm Ring} (\chi, \theta)$ and $\langle \delta_{\rm 3D, m}\rangle_{\rm Ring}(\chi, \theta)$.

The measurements are shown in Fig.~\ref{fig: Buzzard_redmagic_sim_3D_stacked_0.35_z_0.5} as data points for $\langle \delta_{\rm 3D, m}\rangle_{\rm Ring}(\chi, \theta)$ and as shaded bands for $\langle \delta_{\rm 3D, g}\rangle_{\rm Ring} (\chi, \theta)$, for three fixed $\theta$ bins. The plot shows the density constrast as a function of the line-of-sight comoving distance, $\chi_r$. 
The error bars are given by jackknife resampling of the void (upper panel) and cluster (lower panel) catalogues.
We see that, at a fixed angular bin, $\theta$, and tomographic bin, the average line-of-sight structure for both voids and clusters follows the lens redshift distribution, $p_{\rm lens}(\chi)$. This is perhaps expected because, on average, we are simply stacking a set of lines of sight which are slightly overdense or underdense compared to the overall density distribution. This is the case for both galaxy and dark matter density.

Hence, we take the following approximation as our toy model:
\begin{equation}
\begin{aligned}
    \langle \delta_{\rm 3D, g} \rangle_{\rm Ring} (\theta, \chi) =  \frac{b\langle \delta_{\rm m} \rangle_{\rm Ring} (\theta)  \, p_{\rm lens}(\chi)}{\int d\chi' \ p^2_{\rm lens}(\chi') },
\end{aligned}
    \label{eq: model}
\end{equation}
where the denominator comes from normalization given Eq.~\ref{eq: dens2}.
The solid lines in Fig.~\ref{fig: Buzzard_redmagic_sim_3D_stacked_0.35_z_0.5} show this toy model for $\langle \delta_{\rm 3D, g} \rangle_{\rm Ring} (\theta, \chi)$, and it has reasonable agreement with the measured data points.

With this toy model, we interpret the functions $F_{ij}(\cosmoparams, \theta)$ as defined in Eq.~\ref{eq: F} and compare the theoretical predictions with that measured using noiseless, true $\kappa$ maps from the Gower Street simulations.
Fig.~\ref{fig: F_cosmo_buzzard} shows the comparison of the noiseless template $\tilde{F}_{ij}(\cosmoparams, \theta)$ at \textit{Planck} cosmology with the theory calculation for various lens-source combinations. 
The black lines are the toy model prediction, and do not have any scale dependence. The yellow solid lines and dashed lines show the simulation measurements for clusters and voids respectively.
There is reasonable agreement between the measurements and the predicted values, although the measurements for clusters seem to have a mild scale dependence and lower amplitude, which are not captured by the simple model. 

\section{Systematic impacts on the analytic toy model}
\label{sec: systematics}

In this section, we use the analytic toy model to assess qualitatively the impact of systematic effects on the measured profiles: $\langle \delta_{\rm g} \rangle_{\rm Ring} (\theta)$ and $\langle \kappa\rangle_{\rm Ring}(\theta)$. We demonstrate, with an explicit example, the order of magnitude of lens magnification, intrinsic alignment, and photometric redshift uncertainties compared to the profile signal and statistical error. 
We use one realization of the Gower Street simulation with a cosmology close to fiducial: $\Omega_{\rm m}=0.290, \Omega_b=0.050, h=0.667, \sigma_8=0.766, w=-1.01$. Note that in the baseline analysis, all of these effects are forward-modelled.

\begin{figure*}
     \begin{subfigure}[b]{\textwidth}
         \centering
         \includegraphics[width=\textwidth]{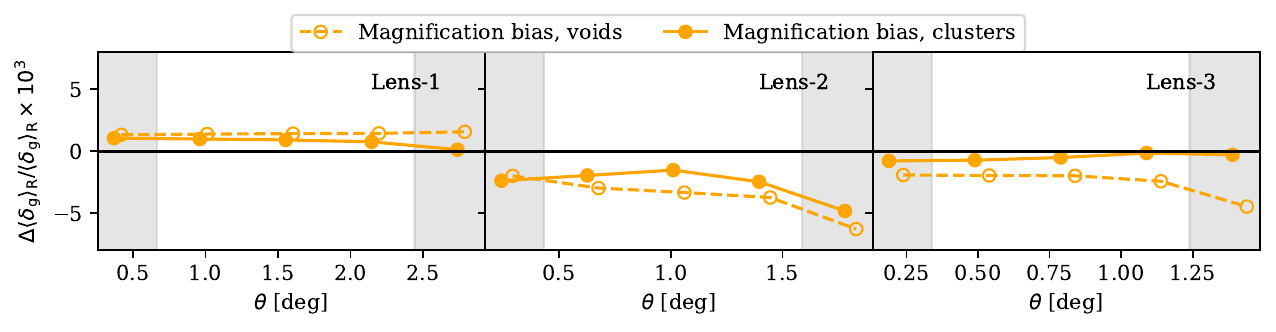}
     \end{subfigure}
    \hfill
    \begin{subfigure}[b]{\textwidth}
         \centering
         \includegraphics[width=\textwidth]{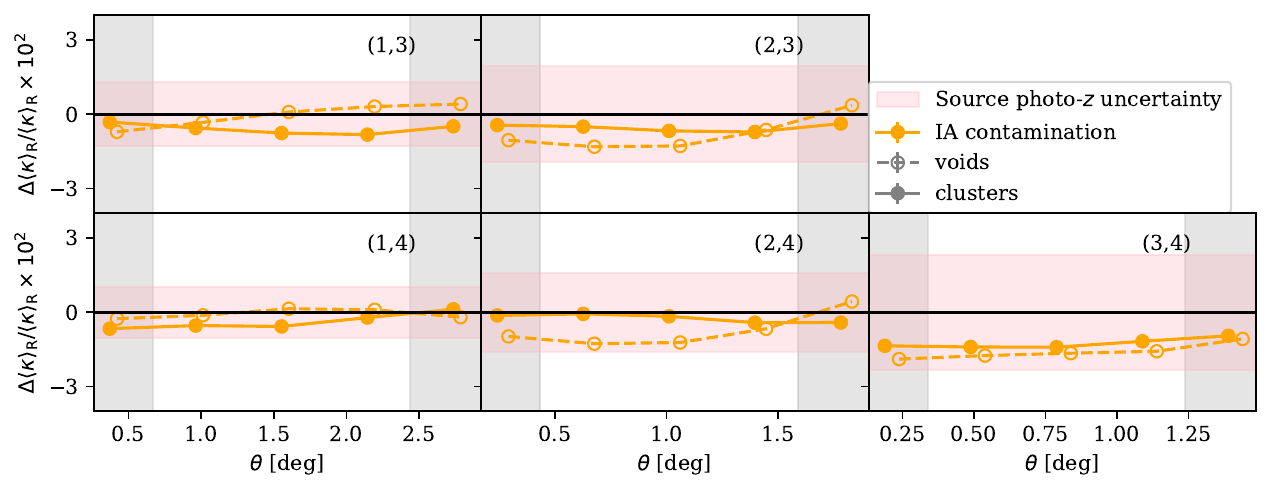}
     \end{subfigure}
        \caption{DES Y3-like weak lensing systematic impacts on the measured profiles using the analytic toy models. Upper panel: Fractional magnification bias contribution on the stacked galaxy density profiles. 
        Lower panel: Fractional intrinsic alignment contamination and source photometric redshift uncertainty on the stacked lensing profiles. The former is shown by connected circles and the latter by pink bands.
        }
        \label{fig: magnification and ia}
\end{figure*}

\subsection{Magnification}
\label{sec: Magnification}

The observed galaxy number density is modulated by weak lensing magnification due to the foreground matter, i.e. $\delta^{\rm obs}_{\rm g}=\delta^{\rm int}_{\rm g}+\delta_{\mu}$. 
The sign of $\delta_{\mu}$ is the result of two competing effects: 1) positive magnification increases the area element at a position $\hat{\mathbf{r}}$ on the sky, diluting the galaxy number density; 2) apparent image size increases, boosting faint galaxies into the photometric sample.
Specifically, the modulation to the galaxy density contrast can be expressed by
\begin{equation}
    \delta_{\mu}(\hat{\mathbf{r}})=2(\alpha-1)\kappa_{\rm lens}(\hat{\mathbf{r}}),
\end{equation}
where $\alpha$ is the response of the number of selected galaxies per unit (unlensed) area to variations in convergence $\kappa$,
and $\kappa_{\rm lens}$ is the convergence experienced by the lens galaxies. For the DES Y3 \texttt{redMaGiC} high-density sample, the slope is measured to be $\alpha=1.31, -0.52, 0.34$ for the three lens redshift bins respectively \citep{Elvin_Poole_2023}.

We produce a magnification map for each lens tomographic bin in the Gower Street simulation; these are then added to the galaxy density map. We rerun the superstructure-finding algorithm and measure the stacked galaxy density profiles at these superstructures.
We find that there is little change in terms of the superstructures found.
The upper panel of Fig.~\ref{fig: magnification and ia} shows the factional difference between the galaxy density profiles with and without magnification.
The impact on $\langle \delta_{\rm g}\rangle_{\rm Ring}(\theta)$ for all tomographic bins is of order $2 \times 10^{-3}$, and the impact decreases with scale.
This is about $1\%$ of the density profile signal and $5\%$ of the statistical error. 

\subsection{Intrinsic Alignment}
\label{sec: IA}

Intrinsic alignment (IA) refers to the effect that the observed galaxy shear contains a non-lensing, intrinsic contribution, i.e. $e^{\rm obs}=e^{\gamma} + e^{\rm IA}$.
IA is caused by the alignment of nearby galaxies (due to tidal fields) producing correlation with the underlying matter field. When the source galaxies have redshift overlap with the lens galaxies, we expect that the reconstructed $\kappa$ profile will contain an IA contribution.
The signal can be modelled in the Non-Linear Alignment \citep[NLA;][]{2004PhRvD..70f3526H, 2007NJPh....9..444B} model framework as:
\begin{equation}
    \kappa_{\rm IA}(\hat{\mathbf{r}}, z)=A_1 (z) \delta_m (\hat{\mathbf{r}}, z),
    \label{eq: ia1}
\end{equation}
where
\begin{equation}
    A_1(z)=-a_1\bar{C}_1 \frac{\rho_{\rm crit}\Omega_{\rm m}}{D(z)}\left( \frac{1+z}{1+z_0} \right)^{\eta_1},
\end{equation}
$\bar{C_1}=5\times10^{-14}M_{\odot}h^{-2}{\rm Mpc}^2$, $z_0=0.62$ is the pivot redshift, $\rho_{\rm crit}$ is the critical density at $z=0$, $D(z)$ is the growth factor normalised to $D(0)=1$, and $a_1$ and $\eta_1$ are free parameters of the model. Finally, for the source catalogue, the IA signal is found by integrating the redshift distribution
\begin{equation}
    \kappa_{\rm IA}^{\rm tot}(\hat{\mathbf{r}})=\int \kappa_{\rm IA} (\hat{\mathbf{r}}, z) \, n_s(z)\, dz.
    \label{eq: ia2}
\end{equation}
The constraints on $a_1$ and $\eta_1$ from the DES Y3 cosmic shear analysis are $a_1=0.36^{+0.43}_{-0.36}$ and $\eta_1=1.66^{+3.26}_{-1.05}$ \citep{2022PhRvD.105b3515S}.

Using the above equations and parameters we compute maps of $\kappa_{\rm IA}$ for each source tomographic bin for the Gower Street simulation. We then use the inverse of Eq.~\ref{eq: KS} to convert these convergences into shear maps, which are added to the true shear maps of the source bin. Finally we convert back to $\kappa$ maps.
The lower panel of Fig.~\ref{fig: magnification and ia} shows the fractional difference between the lensing profiles with and without IA. 
The impact is $<3\%$ on the $\langle \kappa\rangle_{\rm Ring}(\theta)$ signal, and is $<10\%$ of the statistical error. As expected, it is most significant in the lens-source combination $(3,4)$.

\subsection{Photometric redshift uncertainty}

We only consider the photometric redshift uncertainty for the source sample, because \texttt{redMaGiC} galaxies have rather accurate photo-$z$ thanks to their characteristic spectral features.
We compute the scatter of the lensing profiles using 200 realizations of the source redshift distributions in each bin, randomly selected from \textsc{HyperRank} samples \citep{2022MNRAS.511.2170C}. 
The lower panel of Fig.~\ref{fig: magnification and ia} shows, as pink bands, the scatter normalized by the signal. We see that this effect is similar in size to the IA effect. 

\section{Scaling relation validation with Gower Street simulations}
\label{apdx: Model validation with Gower Street simulations}

\begin{figure*}
         \centering
         \includegraphics[width=\textwidth]{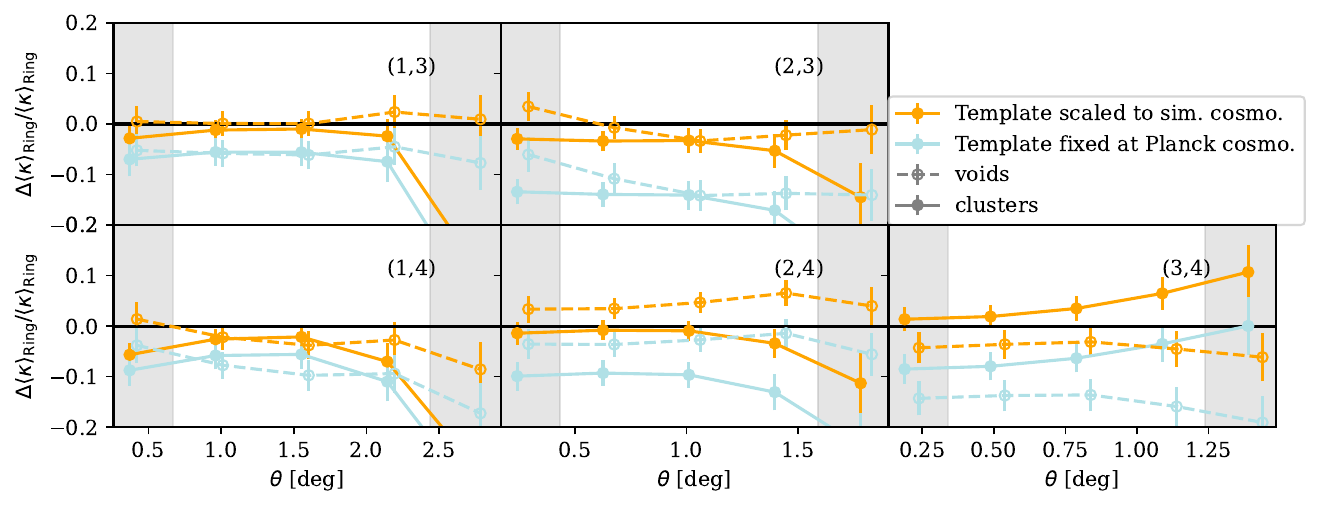}
        \caption{Validation test for the cosmological dependence of $F_{ij}(\Omega_{\rm m}, w, \theta)$ (Eq.~\ref{eq: F scale}) using 103 test simulations from the Gower Street suite. 
        The $y$-axis shows the fractional difference in $\langle \kappa \rangle_{\rm Ring}(\theta)$ from the template model and measurements. The orange lines show the case where the template is scaled to the respective cosmologies of each simulation, and the blue lines show that fixed at \textit{Planck} cosmology.
        Voids and clusters are denoted by dashed and solid lines, respectively.
        }
       \label{fig: scale test}
\end{figure*}

We validate the cosmological scaling relation in Eq.~\ref{eq: F scale} and the determination of the scaling parameter using a `test' sample of 103 Gower Street simulations. The test sample is a random subsample of the total sample, hence has similar cosmological parameter distributions. 
We take the template $\tilde{F}_{ij}(\theta)$ using the training sample at \textit{Planck} cosmology, and scale it to each test sample cosmology before applying Eq.~\ref{eq: F} to compute a model for the lensing profile. 
The difference $\Delta \langle \kappa\rangle_{\rm Ring}(\theta)$ between the template model and measurement is measured for each simulation in the test sample. 
Fig.~\ref{fig: scale test} shows the mean and scatter for $\Delta \langle \kappa\rangle_{\rm Ring}(\theta)$, with and without scaling out the cosmology, normalized by the average lensing profiles, $\langle \kappa\rangle_{\rm Ring}(\theta)$, of the test samples. 
The unscaled samples, i.e. using the templates at $\textit{Planck}$ cosmology, can systematically bias the predicted $\kappa$ profiles by about $10\%$. 

\section{Likelihood construction}
\label{apdx: likelihood}

We begin with the general Gaussian likelihood with both noise contributions via covariances  $\mathbfss{C}_{\kappa}$ and $\mathbfss{C}_{g}$:
\begin{equation}
\begin{aligned}
\mathcal{L}(\hat{b}_{\rm prof}\,)\;\propto & \;\exp\Biggl[-\tfrac{1}{2}\,\Bigl(
\hat{b}_{\rm prof}\langle \kappa_{j}\rangle^{i}_{\rm Ring}(\theta)
\;-\;
F_{ij}(\cosmoparams, \theta)\,\langle \delta_{\rm g} \rangle^{i}_{\rm Ring}(\theta)
\Bigr)^{T}
\\
& \times \bigl(\hat{b}_{\rm prof}^{2}\mathbfss{C}_{\kappa} \;+\; \mathbfss{C}_{g}\bigr)^{-1}
\\
& \times \Bigl(
\hat{b}_{\rm prof}\langle \kappa_{j}\rangle^{i}_{\rm Ring}(\theta)
\;-\;
F_{ij}(\cosmoparams, \theta)\,\langle \delta_{\rm g} \rangle^{i}_{\rm Ring}(\theta)
\Bigr)\Biggr].
\end{aligned}
\end{equation}
\noindent Note that here the likelihood begins in in terms of $\hat{b}_{\rm prof}$ (rather than explicitly $1/\hat{b}_{\rm prof}$); we have in no way chosen $1/\hat{b}_{\rm prof}$ as being special.

We assume that the noise in our estimate of the covariance is dominated by the weak lensing contribution, roughly $\mathbfss{C}_{\kappa}\gg \mathbfss{C}_{g}$, so that $\bigl(\hat{b}_{\rm prof}^{2}\mathbfss{C}_{\kappa} + \mathbfss{C}_{g}\bigr)^{-1}
\;\approx\;
\frac{1}{\hat{b}_{\rm prof}^{2}} \mathbfss{C}^{-1}$, in which $\mathbfss{C}$ is not bias-dependent. Therefore, up to an overall constant,
\begin{equation}
\begin{aligned}
\mathcal{L}(\hat{b}_{\rm prof})\;\propto & \;\exp\Bigl[
-\tfrac{1}{2}\,
\Bigl(
\langle \kappa_{j}\rangle^{i}_{\rm Ring}(\theta)
\;-\;
\tfrac{1}{\hat{b}_{\rm prof}}\,F_{ij}(\cosmoparams, \theta)\,\langle \delta_{\rm g} \rangle^{i}_{\rm Ring}(\theta)
\Bigr)^{T} \\
& \mathbfss{C}^{-1}\,
\Bigl(
\langle \kappa_{j}\rangle^{i}_{\rm Ring}(\theta)
\;-\;
\tfrac{1}{\hat{b}_{\rm prof}}\,F_{ij}(\cosmoparams, \theta)\,\langle \delta_{\rm g} \rangle^{i}_{\rm Ring}(\theta)
\Bigr)
\Bigr].
\end{aligned}
\end{equation}

\begin{figure*}
	\includegraphics[width=\textwidth]{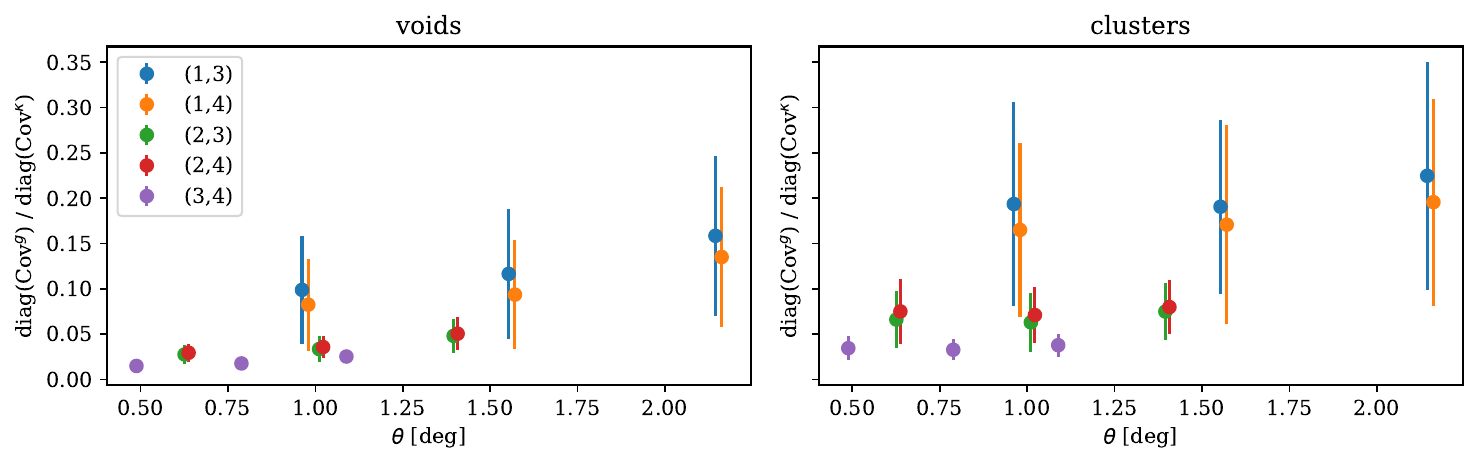}
    \caption{Ratio of the diagonal terms of $\hat{b}^2_{\rm prof}\mathbfss{C}_{\kappa}$, where $\hat{b}^2_{\rm prof}=1.7$, and $\mathbfss{C}_{g}$, using Jackknife resampling measured from the Gower Street simulations. The lens - source combination is denoted in the legend, and the $\theta_{3,4}$ denotes the angular scales used }
    \label{fig: diag_cov_ratio.pdf}
\end{figure*}

Figure~\ref{fig: diag_cov_ratio.pdf} shows the comparison of the diagonal terms of $\hat{b}^2_{\rm prof}\mathbfss{C}_{\kappa}$, where $\hat{b}^2_{\rm prof}=1.7$, and $\mathbfss{C}_{g}$, using Jackknife resampling measured from the Gower Street simulations. We can see that for lens bin 2 and 3, the ratio of the two covariances is below $0.1$ for this nominal galaxy bias value. For bin 1, the covariance of $\mathbfss{C}_{g}$ is larger, at about $0.15-0.2$. This can be partially attributed to the small sample size of the voids and clusters, as can also been seem in the large scatter between simulations.

\section{Correlation matrix}
\label{apdx: covariance}

As described in Section~\ref{sec: Covariance from the Gower Street simulations}, we compute covariance matrices for the data vectors using Eq.~\ref{eq: cov} from the 415 Gower Street mocks. For each mock, we compute the noise of the $\kappa$-profiles as the difference between the measured profile and the theory profile, where the theory is produced using the template scaled to the mock cosmology, and with fixed galaxy bias value, $\hat{b}_{\rm prof} = 1.7$. The covariance matrix is then computed as the average outer product of these noise vectors. Data points in the same lens bin are concatenated.

\begin{figure*}
	\includegraphics[width=\textwidth]{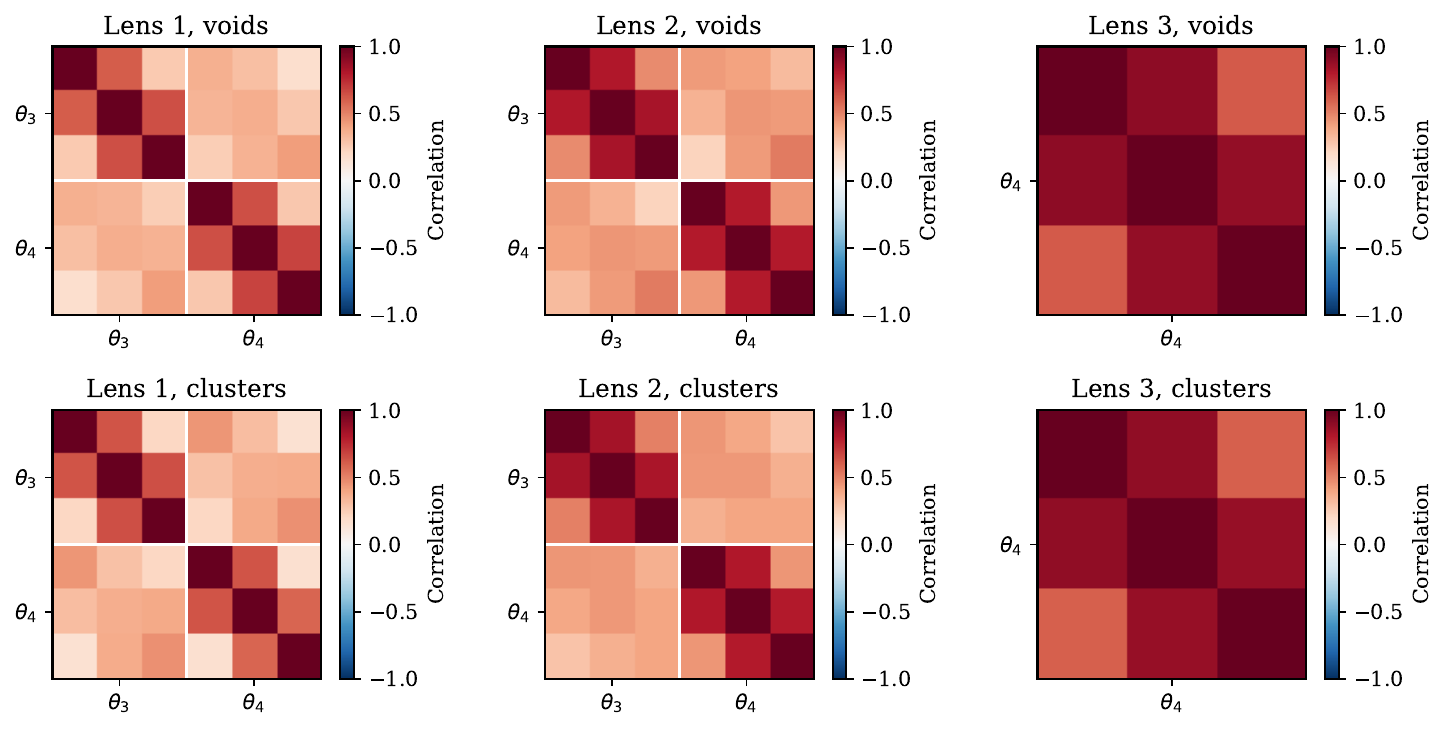}
    \caption{Correlation matrices for the lensing profiles, $\langle \kappa\rangle_{\rm Ring}(\theta)$, computed using 415 simulations from the Gower Street suite. Each lens bin contains different data vector lengths due to the different lens-source combination adopted. The $\kappa$-profile measured with the $i$-th source bin is denoted by the subscript of $\theta$ on the axis of each figure.}
    \label{fig: Dirac-correlation-source-split}
\end{figure*}

Figure~\ref{fig: Dirac-correlation-source-split} illustrates the correlation structure of the data vectors, $\langle \kappa\rangle_{\rm Ring}(\theta)$, for the three lens tomographic bins. The correlation matrix is defined as ${\rm Corr}_{ij} = C_{ij}/\sqrt{C_{ii}C_{jj}}$, where $C_{ij}$ is the covariance matrix. The angular scales are abbreviated to $\theta_i$ for the mass map in bin $i$.

The significant off-diagonal terms in the correlation matrices indicate that the profile measurements are heavily correlated.

\section{Superstructure size}
\label{apdx: superstructure size}

\begin{figure*}
	\includegraphics[width=0.7\textwidth]{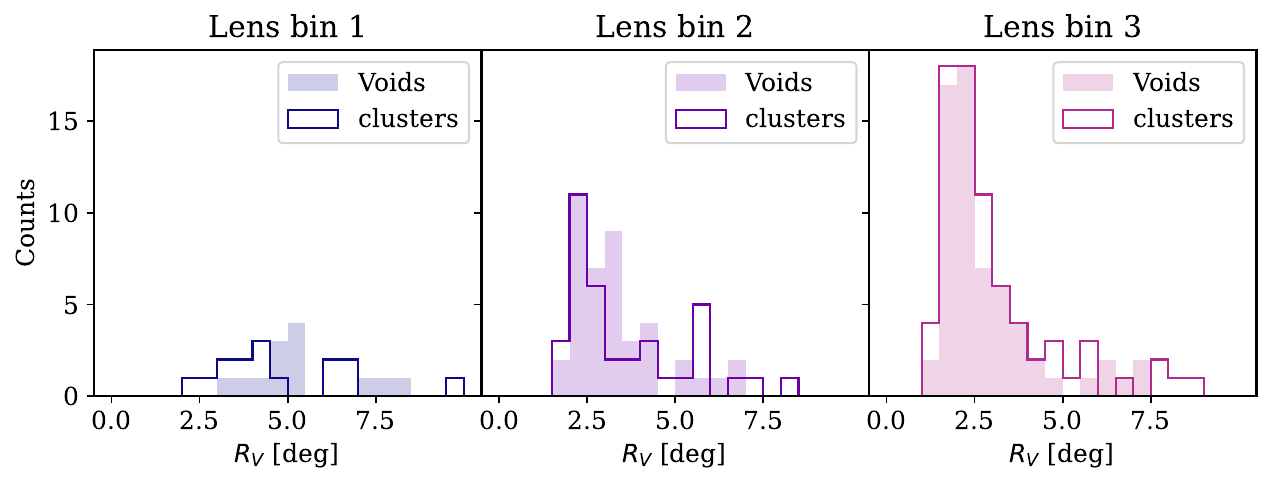}
    \caption{Distribution of the radii, $R_V$, for the identified voids (shaded region) and clusters (solid lines) in lens bin 1 - 3 for the DES Y3 RedMaGiC galaxies.}
    \label{fig: superstructure-radius}
\end{figure*}

In the structure finder algorithm described in Section~\ref{sec: algo}, we define the angular size of the voids and clusters $R_V$. 
Fig.~\ref{fig: superstructure-radius} shows the distribution of these superstructure radii found in lens bins 1 - 3 in degrees. We see that most objects have a size about twice as large as the smoothing scale, and the voids and clusters have a similar size distribution. 

Although we do not use $R_V$ in our main analyses, we note that one can in principle measure the profile as a function of the scaled size, i.e. $\theta/R_V$. This is adopted in similar analysis in the literature \citep[e.g.][]{2021MNRAS.505.4626J}, to account for the different sizes in the stacked profiles, assuming that they are self-similar. This is not done in this analysis as we do not include any scaling in Eq.~\ref{eq: F}, and we leave this point for future exploration.

\section{Cuts in the mask}
\label{apdx: Cuts in the mask}

In Section~\ref{sec: redmagic}, we introduced an additional mask to remove the residual density gradient at $70^{\circ}<{\rm RA}<330^{\circ}$. This corresponds to removing about $20\%$ of the original footprint. In this section, we demonstrate the impact of this cut on the measured DES Y3 data vector. 

\begin{figure*}
     \begin{subfigure}[b]{\textwidth}
         \centering
         \includegraphics[width=\textwidth]{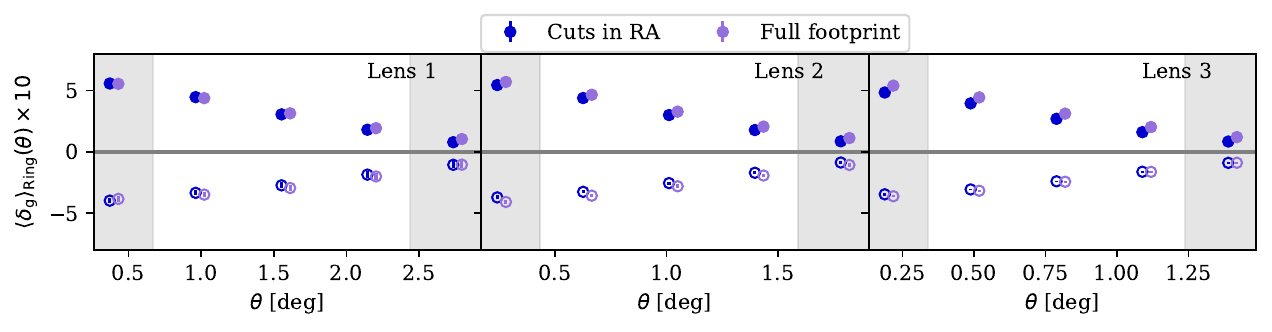}
     \end{subfigure}
    \hfill
    \begin{subfigure}[b]{\textwidth}
         \centering
         \includegraphics[width=\textwidth]{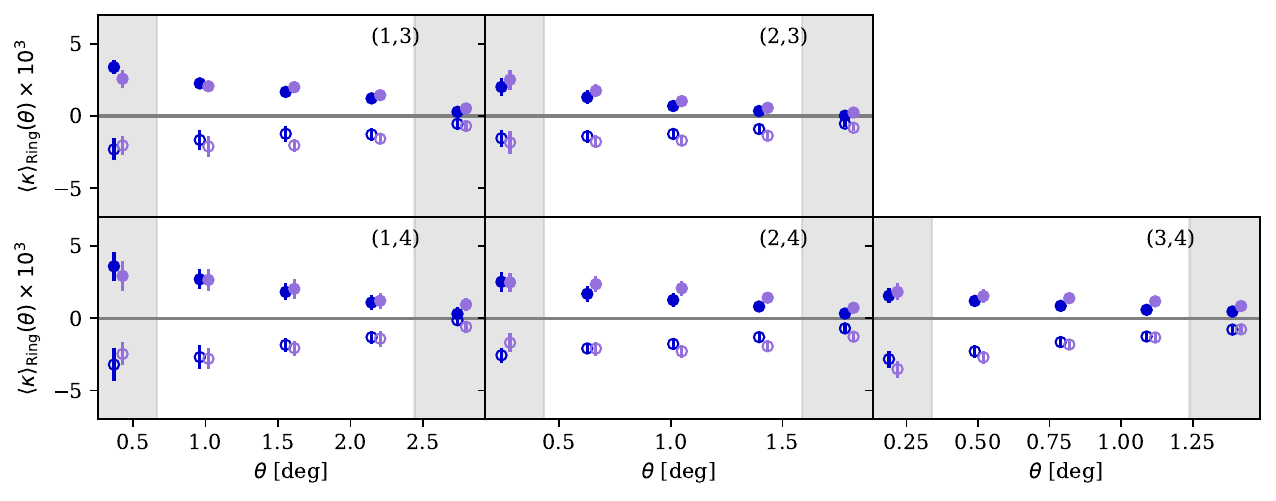}
     \end{subfigure}
        \caption{The impact of the mask cut on the measured galaxy density and kappa profiles. The fiducial sample used in the main analysis with cuts is shown in dark blue, and the sample without cuts is shown in purple. Data points are shifted horizontally for visual clarity.
        }
        \label{fig: full vs racut}
\end{figure*}

We apply the structure-finding algorithm on the galaxy density contrast map with the original DES Y3 mask to obtain a new catalogue of voids and clusters. We then measure the stacked galaxy density contrast and mass map profiles for this new catalogue. 
Fig~\ref{fig: full vs racut} shows the profiles measured with the original mask (`full footprint'), compared to the main analysis (`cuts in RA'). 
There are visible shifts in the profiles measured using the two different catalogues. The shift is slightly more prominent for clusters in lens bins 2 and 3, and the direction is coherent for both the galaxy density and $\kappa$-profiles. 
This means that, if we were to adopt the same template model on the full footprint, it is likely that the measured galaxy bias will be consistent with the main analysis.

\bsp	
\label{lastpage}
\end{document}